\documentclass[12pt,preprint]{aastex}

\newcommand{\kms}{km~s$^{-1}$}
\newcommand{\ie}{{\it i.e., }}
\newcommand{\etal}{{\it et al. }}
\newcommand{\eg}{{\it e.g., }} 

\shorttitle{SN 1993J}
\shortauthors{Weiler, \etal}

\begin{document}

\title{Long Term Radio Monitoring of SN~1993J}

\author{Kurt W.~Weiler}
\affil{Naval Research Laboratory, Code 7210, 
Washington, DC 20375-5351; Kurt.Weiler@nrl.navy.mil}

\author{Christopher L.~Williams\altaffilmark{1}}
\affil{Naval Research Laboratory, Code 7213, 
Washington, DC 20375-5351; clmw@mit.edu}

\author{Nino Panagia\altaffilmark{2,}\altaffilmark{3}}
\affil{Space Telescope Science Institute, 3700 San Martin Drive, Baltimore, MD 
21218; panagia@stsci.edu}

\author{Christopher J.~Stockdale and Matthew T. Kelley}
\affil{Marquette University, Physics Department, P.O. Box 1881, Milwaukee, WI 53214-1881; christopher.stockdale@mu.edu and matthew.kelley@mu.edu}

\author{Richard A.~Sramek}
\affil{ PO Box 0, National Radio Astronomy Observatory, Socorro, NM 87801; dsramek@nrao.edu}

\author{Schuyler D.~Van Dyk}
\affil{IPAC/Caltech, Mail Code 100-22, Pasadena, CA 91125; vandyk@ipac.caltech.edu}

\and

\author{J.M.~Marcaide}
\affil{Departamento de Astronomia, Universitat de Valencia, 46100 Burjassot, Spain; J.M.Marcaide@uv.es}

\altaffiltext{1}{Present address: Massachusetts Institute of Technology,
Kavli Institute for Astrophysics and Space Research, Cambridge, MA
02139}
\altaffiltext{2}{INAF - Osservatorio Astrofisico de Catania, Via S. Sofia 78, I-95123 Catania, Italy}
\altaffiltext{3}{Supernova Ltd., Olde Yard Village \#131, Northsound
Road, Virgin Gorda, British Virgin Islands.}

\begin{abstract}
We present our extensive observations of the radio emission from
supernova (SN) 1993J, in M~81 (NGC~3031), made with the Very Large Array,
at 90, 20, 6, 3.6, 2, 1.2, and 0.7 cm, as well as numerous
measurements from other telescopes and at other wavelengths. The combined data set constitutes probably the most detailed set of measurements ever established for any
SN outside of the Local Group in any wavelength range. Only the very subluminous SN~1987A in the Large Magellanic Cloud has been the subject of such an intensive observational program.  The radio emission evolves regularly in both time and frequency, and the usual interpretation in terms of shock interaction
with a circumstellar medium (CSM) formed by a pre-supernova stellar wind describes the observations rather well considering the complexity of the phenomenon.  However: 1) The
highest frequency measurements at 85 - 110 GHz at early times ($<40$
days) are not well fitted by the parameterization which describes the cm
wavelength measurements rather well. 2) At mid-cm wavelengths there is
often deviation from the fitted radio light curves, particularly near
the peak flux density, and considerable shorter term deviations in the
declining portion when the emission has become optically thin. 3) At a time
$\sim3100$ days after shock breakout, the decline rate of the
radio emission steepens from (t$^{+\beta}$) $\beta \sim -0.7$ to $\beta
\sim -2.7$ without change in the spectral index ($\nu^{+\alpha}$; $\alpha \sim -0.81$). However, this decline is best described not as a power-law, but as an exponential decay starting at day 3100 with an
e-folding time of $\sim 1100$ days. 4) The best overall fit to all of the data is a model including both non-thermal synchrotron self-absorption (SSA) and a thermal free-free absorbing
(FFA) components at early times, evolving to a constant spectral index, optically thin decline rate, until a break in that decline rate at day $\sim3100$ as mentioned above. Moreover, neither a
purely SSA nor a purely FFA absorbing models can provide a  fit that
simultaneously reproduces the light curves, the spectral index evolution,
and the brightness temperature evolution. 5) The radio and X-ray light
curves display quite similar behavior and both suggest a sudden drop in the supernova progenitor mass-loss rate at $\sim 8000$ years prior to shock breakout.
\end{abstract}


\section{Introduction}

SN~1993J [RA(J2000) = $9^{\rm h}55^{\rm m}24{\fs}7740 \pm 0{\fs} 0006$, 
Dec(J2000) =   +$69^{\circ}01\arcmin 13{\farcs}700 \pm 0\farcs 003$;
\citealt{Marcaide93a}] in M 81 (NGC 3031) was discovered at magnitude
$V\sim11{\fm}8$ on 28.91 March 1993 \citep{Ripero93}, and by 30 March,
at maximum optical magnitude $V = 10{\fm}7$, had become the brightest supernova
(SN) in the northern hemisphere since SN 1954A.  Hydrogen was soon
identified in its optical spectrum, classifying it as a type II SN
(SNII) \citep[see, e.g., ][]{Gomez93,Andrillat93,Filippenko93a}.

From the outset, SN~1993J displayed unusual characteristics for a
SNII.    The visual light curve was markedly different from both the
SNIIL (linear) and SNIIP (plateau) subtypes in that it exhibited a
second maximum $\sim17$ days after the first one \citep{VanDriel93}.  Its
unusual light curve and spectrum were quickly interpreted by
\citet{Podsiadlowski93,Nomoto93,Swartz93} as implying a red supergiant
progenitor with a thin hydrogen envelope which would spectrally evolve
from resembling a SNII to resembling a SNIb, thereby suggesting a SNIIb
classification for SN 1993J.  Continuing observations of visual spectra
by \citet{Filippenko93b} confirmed this transition.

Due to its proximity ($3.63 \pm 0.34$ Mpc; \citealt{Freedman94}) and the
fact that  SNII are expected to be strong radio emitters
\citep{Weiler89}, \citet{Sramek93} made  very early attempts with the
Very Large Array (VLA)\footnote{The VLA telescope of the National Radio
Astronomy Observatory is operated by Associated Universities, Inc. under
a cooperative agreement with the National Science Foundation.} to detect
the SN.  After establishing upper limits at 3.6 cm and 20 cm on  UT
31.07 March 1993 \citep{Sramek93}, radio emission was detected with the
VLA on UT 02.30 April 1993, with a flux density of $0.8 \pm 0.2$ mJy at
1.3 cm (\citealt{Weiler93}, see also \citealt{VanDyk93a,VanDyk93b}) and
with the Ryle Telescope in Cambridge, UK on UT 5.7 April 1993 at 2 cm
(15.3 GHz) by \citet{Pooley93a}.  By UT 25 April 1993 the rapidly expanding
SN already had a measurable size of $0.25 \pm 0.1$ milliarcseconds (mas)
with Very Long Baseline Interferometry (VLBI) techniques (\citealt{Marcaide93a,Marcaide93b}).

High angular resolution VLBI size measurements of the expanding
SN were conducted very early by \citet{Marcaide94} and \citet{Bartel94}
and VLBI monitoring continues to the present at multiple wavelengths (see, e.g.,
\citealp{Marcaide95a,Marcaide95b,Marcaide97,Marcaide05,Marcaide07,Bartel00,Bartel02,Bietenholz01,Bietenholz03}).

Extensive radio monitoring of the integrated flux density of SN~1993J
has been conducted by the VLA at 20 cm (1.4 GHz), 6 cm (4.9 GHz), 3.6 cm
(8.4 GHz), 2 cm (14.9 GHz) and 1.2 cm (22.5 GHz) \citep{VanDyk94} and
with the Ryle Telescope at 2 cm (15.3 GHz) \citep{Pooley93b}. 
Additional observations were also conducted at 0.3 cm (85 - 110 GHz)
with the Institut de Radioastronomie Millimetrique (IRAM) telescope \citep{Radford93} and at the
Caltech Owens Valley Radio Observatory (OVRO)
\citep{Phillips93a,Phillips93b}, and at 0.9 cm with the Effelsberg 100-m
telescope of MPIfR (W.~Reich, private communication).  More recently,
\citet{Chandra01} have conducted observations with the Giant Metrewave Radio Telescope (GMRT) in India at 49 cm
(0.6 GHz) and 20 cm (1.4 GHz) and we have added new measurements with
the VLA at 90 cm (0.3 GHz) and 0.7 cm (43 GHz).  

In this paper we consider the integrated flux density measurements and
their physical interpretation.

\section{Radio Observations}\label{observations}

Almost a decade and a half has passed since the explosion of SN 1993J so
that it is an appropriate interval to consider the extensive set of
radio observations which are now available from 90 cm at the longest
wavelength to 0.3 cm at the shortest.  In addition to previously
published VLA results \citep{VanDyk94}, we present here almost 200 new
VLA observations of SN~1993J at 90, 20, 6, 3.6, 2, 1.2, and 0.7 cm along
with all published results which could be found in the literature or
have been provided to us as private communications at 49, 20, 0.9, and 0.3
cm.  

All of the available data are presented in Tables \ref{tab1} and
\ref{tab2} and the data at the best sampled wavelengths of 90, 49, 20,
6, 3.6, 2, 1.2 and 0.3 cm, principally from the VLA and the Cambridge
Ryle telescope, along with contributions from the IRAM and OVRO millimeter telescopes and the GMRT, are plotted in and Figures \ref{fig1}, \ref{fig2}, \ref{fig8}, \ref{fig11}, and \ref{fig13}.  The previously published results from \citet{VanDyk94} are also included in Tables \ref{tab1} and \ref{tab2} and plotted in the Figures for completeness and ease of reference.  However, to reduce the size and complexity of Figures \ref{fig1}, \ref{fig2}, \ref{fig8}, \ref{fig11}, and \ref{fig13}, the sparse measurements at 32 and 43 GHz are not plotted
even though they were used in the fitting procedure.

The techniques of observation, editing, calibration, and error
estimation are described in previous publications on the radio emission
from SNe \citep[see, e.g., ][]{Weiler86,Weiler90}.  The ``primary''
calibrator was 3C286, which is assumed to be constant in time with flux
densities of 25.84, 14.45, 7.42, 5.20, 3.45, and 2.52 Jy at 90, 20, 6,
3.6, 2, and 1.2 cm, respectively.  The ``secondary''
calibrator\footnote{Secondary calibrators are chosen to be compact and
unresolved by the longest VLA baselines.  While compact and serving as
good phase references, such objects are usually variable, so that their
flux density must be recalibrated regularly from the primary calibrators.} was normally J1048+717\footnote{Several of the early
observations at 90 cm used J0834+555, with a position of RA(J2000) =
$08^{\rm h}34^{\rm m}54{\fs}904117$, Dec(J2000) = +$55^{\circ}34\arcmin
21{\farcs}070980$ as a secondary calibrator. Also, between 11 September 1993 and 08 February 1994, J0949+662 [RA(J2000) = $09^{\rm h}49^{\rm m}12{\fs}2100$, Dec(J2000) = +$66^{\circ}14\arcmin
59{\farcs}321$] was used as a secondary calibrator at 20 cm.}, with a defined position
of RA(J2000) = $10^{\rm h}48^{\rm m}27{\fs}619917$, Dec(J2000) =  
+$71^{\circ}43\arcmin 35{\farcs}938280$.  After flux density calibration
by 3C286, it served as the actual gain and phase calibrator for
SN~1993J.  As expected for secondary calibrators, the flux density of
J1048+717 has been varying over the years, as can be seen in Table
\ref{tab3} and Figure \ref{fig3}.

The flux density measurement errors for SN~1993J are a combination of
the rms map error, which measures the contribution of small unresolved
fluctuations in the background emission and random map fluctuations due
to receiver noise, and a basic fractional error $\epsilon$, included to
account for the normal inaccuracy of VLA flux density calibration (see,
e.g., \citealt{Weiler86}) and possible deviations of the primary
calibrator from an absolute flux density scale.  The final errors
($\sigma_f$) given for the measurements of SN~1993J are taken as

\begin{equation}
\label{eq1}
\sigma_{f}^{2} = (\epsilon S_0)^2+\sigma_{0}^2 
%\label{eq:err}
\end{equation} 

\noindent where $S_0$ is the measured flux density, $\sigma_0$ is the
map rms for each observation, and $\epsilon = 0.15$ for 90 cm, 0.10 for
20 cm, 0.05 for 6 and 3.6 cm, 0.075 for 2 cm, and 0.10 for 1.2 cm. All
upper limits are listed as three sigma ($3\sigma$). 

The appropriate errors to use for the Cambridge measurements at 2 cm are difficult to determine. The authors \citep{Pooley93b} mention that the variable nucleus of M81 is not fully resolved from SN~1993J, and that their calibrator B0954+658 is clearly variable. However, they have done their best to remove such effects and estimate that 5\% ``is a good estimate of the uncertainty in the observations.'' Because of our knowledge of the uncertainties of observations with the VLA at 2 cm wavelength, and our decision to use a standard minimum error of 7.5\% in Equation \ref{eq1} at that wavelength, we have chosen to assign a 10\% error to all Cambridge data to additionally account for possible systematic effects between the two telescopes and the two different secondary calibrators. Such a value may well be too conservative but, because of the large number of multi-frequency points in the data set, the assumption of possibility too large errors for the Cambridge data does not affect any of our fits or conclusions.

\section{Radio Supernova Models}

All known RSNe appear to share common properties of: 1) nonthermal
synchrotron emission with high brightness temperature; 2) a decrease in
absorption with time, resulting in a smooth turn-on first at shorter
wavelengths and later at longer wavelengths; 3) a power-law decline of
the flux density with time at each wavelength after the source becomes optically thin at that wavelength; and 4) a final, asymptotic approach of spectral index $\alpha$ ($S \propto
\nu^{+\alpha}$) to an optically thin, nonthermal, constant negative
value \citep{Weiler86,Weiler90}. 

\citet{Chevalier82a,Chevalier82b} proposed that the relativistic
electrons and enhanced magnetic field necessary for synchrotron emission
arise from the SN blastwave interacting with a relatively high density
CSM which has been ionized and heated by the initial UV/X-ray flash. 
This CSM is presumed to have been established by a constant mass-loss
($\dot M$) rate, constant velocity ($w_{\rm wind}$) wind (\ie $\rho
\propto \frac{\dot M}{w_{\rm wind}~r^2}$) from a massive stellar
progenitor or a companion.  This ionized CSM is the source of some or all
of the initial free-free absorption (FFA) although more recently \citet{Chevalier98} has proposed that
synchrotron self-absorption (SSA) may play a role at some times and in some objects.  

A rapid rise in the observed radio flux density results from a decrease in these absorption processes as the radio emitting region expands and
the absorption processes, either internal or along the line-of-sight,
decrease.  \citet{Weiler90} have suggested that this CSM can be
``clumpy'' or ``filamentary,'' leading to a slower radio turn-on, and
\citet{Montes97} have found at least one example for the presence of
a distant ionized medium along the line-of-sight which is time
independent and can cause a spectral turn-over at low radio
frequencies.  In addition to clumps or filaments, the CSM may be
structured with significant density irregularities such as rings, disks,
shells, or gradients and many, if not most, well studied RSNe appear to
show a transition to a significantly less dense CSM after a number of
years (several thousand years in the time frame of the presupernova
wind; see, e.g., SN~1980K, \citet{Montes98}, SN~1988Z, \citet{VanDyk93c,Williams02}, and SN~2001gd, Stockdale
\etal in press).

\subsection{Radio Light Curves}

Following the most recent RSN modeling discussion of \citet{Weiler02}
and \citet{Sramek03}, we adopt a parameterized model :

\begin{equation}
\label{eq2}
S(\mbox{mJy}) = K_1 \left(\frac{\nu}{\mbox{5\
GHz}}\right)^{\alpha} \left(\frac{t-t_0}{\mbox{1\ day}}\right)^{\beta}
e^{-\tau_{\rm external}} \left(\frac{1-e^{-\tau_{{\rm CSM}_{\rm clumps}}}}{\tau_{{\rm CSM}_{\rm
clumps}}}\right) \left(\frac{1-e^{-\tau_{\rm internal}}}{\tau_{\rm internal}}\right) 
\end{equation} 

\noindent with  

\begin{equation}
\label{eq3}
\tau_{\rm external}  =  \tau_{{\rm CSM}_{\rm homogeneous}}+\tau_{\rm distant},
\end{equation}

\noindent where

\begin{equation}
\label{eq4}
\tau_{{\rm CSM}_{\rm homogeneous}}  =  K_2
\left(\frac{\nu}{\mbox{5 GHz}}\right)^{-2.1}
\left(\frac{t-t_0}{\mbox{1\ day}}\right)^{\delta}
\end{equation}

\begin{equation}
\label{eq5}
\tau_{\rm distant}  =  K_4  \left(\frac{\nu}{\mbox{5\
GHz}}\right)^{-2.1}
\end{equation} 

\noindent and

\begin{equation}
\label{eq6}
\tau_{{\rm CSM}_{\rm clumps}}  =  K_3 \left(\frac{\nu}{\mbox{5\
GHz}}\right)^{-2.1} \left(\frac{t-t_0}{\mbox{1\
day}}\right)^{\delta^{\prime}}
\end{equation} 

\noindent with $K_1$, $K_2$, $K_3$, and $K_4$ determined from fits to the data and corresponding, formally, to the flux density ($K_1$), homogeneous ($K_2$, $K_4$), and clumpy or filamentary ($K_3$) FFA 
at 5~GHz one day after the explosion date $t_0$.  The terms $\tau_{{\rm CSM}_{\rm homogeneous}}$ and  $\tau_{{\rm CSM}_{\rm clumps}}$ describe the attenuation of local, homogeneous free-free absorption CSM and clumpy or filamentary free-free absorbing CSM, respectively, that are near enough to the SN progenitor that they are altered by the rapidly expanding SN blastwave.  The $\tau_{{\rm CSM}_{\rm homogeneous}}$ FFA is produced by an ionized medium that completely covers the emitting source (``homogeneous external absorption''), and the $(1-e^{-\tau_{{\rm CSM}_{\rm clumps}}}) \tau_{{\rm CSM}_{\rm clumps}}^{-1}$ term describes the attenuation produced by an inhomogeneous FFA medium \citep[``clumpy absorption''; see][for a more detailed discussion of attenuation in inhomogeneous media]{Natta84}. The $\tau_{\rm distant}$ term describes the attenuation produced by a homogeneous FFA medium which completely covers the source but is so far from the SN progenitor that it is not affected by the expanding SN blastwave and is consequently constant in time.  All external and clumpy absorbing media are assumed to be purely thermal, singly ionized gas which absorbs via free-free absorption (FFA) with frequency dependence $\nu^{-2.1}$ in the radio.  The parameters $\delta$
and $\delta'$ describe the time dependence of the optical depths for the local homogeneous and clumpy or filamentary media, respectively. 

Since it is physically realistic and may be needed in some RSNe where radio observations have been obtained at early times and high frequencies, Equation (\ref{eq2}) also includes the possibility for an internal absorption term\footnote{Note that for simplicity an internal absorber attenuation of the form $\left(\frac{1-e^{-\tau_{{\rm CSM}_{\rm internal}}}}{\tau_{{\rm CSM}_{\rm internal}}}\right)$, which is appropriate for a plane-parallel geometry, is used instead of the more complicated expression \citep[e.g., ][]{Osterbrock74} valid for the spherical case.  The assumption does not affect the quality of the analysis because, to within 5\% accuracy, the optical depth obtained with the spherical case formula is simply three-fourths of that obtained with the plane-parallel slab formula.}.  This internal absorption  ($\tau_{\rm internal}$) term may consist of two parts -- synchrotron self-absorption (SSA; $\tau_{{\rm internal}_{\rm SSA}}$), and mixed, thermal FFA/non-thermal emission ($\tau_{{\rm internal}_{\rm FFA}}$). 

\begin{equation}
\label{eq9}
\tau_{\rm internal}  = \tau_{\rm internal_{\rm SSA}} + \tau_{\rm internal_{\rm FFA}}
\end{equation}

\begin{equation}
\label{eq10}
\tau_{\rm internal_{\rm SSA}} = K_5\left(\frac{\nu}{\mbox{5\
GHz}}\right)^{\alpha-2.5}  \left(\frac{t-t_0}{\mbox{1\
day}}\right)^{\delta^{\prime\prime}}
\end{equation}

\begin{equation}
\label{eq11}
\tau_{\rm internal_{\rm FFA}}  =   K_6  \left(\frac{\nu}{\mbox{5\
GHz}}\right)^{-2.1} \left(\frac{t-t_0}{\mbox{1\
day}}\right)^{\delta^{\prime\prime\prime}}
\end{equation}

\noindent with $K_5$ corresponding, formally, to the internal,
non-thermal ($\nu^{\alpha - 2.5}$) SSA and $K_6$, corresponding, formally,
to the internal thermal ($\nu^{-2.1}$) free-free absorption mixed with
non-thermal emission, at 5~GHz one day after the explosion date $t_0$. 
The parameters $\delta^{\prime \prime}$ and $\delta^{\prime \prime
\prime}$ describe the time dependence of the optical depths for the SSA
and FFA internal absorption components, respectively. 

Application of this basic parameterization has been shown to be effective
in describing the physical characteristics of the presupernova system,
its CSM, and its final stages of evolution before explosion for objects
ranging from the two decades of monitoring the complex radio emission
from SN 1979C \citep{Montes00} through the unusual SN 1998bw (GRB980425)
\citep{Weiler01} and most recent $\gamma$-ray bursters
\citep{Weiler02,Weiler03}.

\subsection{Brightness Temperature}

Given the measured fluxes, we can compute the corresponding brightness
temperatures if the angular size of the radio region is known.  \cite{Marcaide95a,Marcaide95b,Marcaide97,Marcaide05,Marcaide07} have measured the apparent expansion of SN 1993J in the radio with a series of VLBI experiments, starting as early
as day 182 and extending up to day 3858.  \cite{Marcaide07} find
that a power law of the form $r \propto t^m$ can provide a good, frequency independent fit to all observations of the angular diameter at 3.6, 6, and 18 cm with $m = 0.845 \pm 0.005$
through day 1500. Although there are no measurements at early epochs,
the remarkably good quality of the fit up to day 1500 justifies our assuming $m = 0.845$ since the epoch of the SN shock breakout.  Under this assumption we can express the angular expansion of SN~1993J as

\begin{equation}
\label{eq12}
r = 6.2 \times (t/1 {\rm day})^{0.845} ~~~ \mu as
\end{equation}

\noindent which gives the radius $r$ of the circle in microarcseconds ($\mu$as) that encompasses half of
the total radio flux density to better than 20\%, assuming isotropic radio emission.  Adopting this
expansion rate, the brightness temperature turns out to be

\begin{equation}
\label{eq13}
T_B = 1.30\times10^{10} (S_\nu(corr)/2) \lambda^2 t^{-1.69} ~{\rm K}
\end{equation}

\noindent where the radio flux density $S_\nu(corr)$ is the observed flux density, corrected for model estimated external free-free absorption, expressed in mJy, the wavelength ($\lambda$) in cm,
and the time (t) in days.  The term $S_\nu/2$ accounts for the fact that the circular area inside
$r$ is defined to include only half of the total flux density.

\section{Fitting Results}

Note that even approximate fitting of the data with our standard models
requires splitting it into two parts: an ``early'' data set from the first radio detection 
through day $\sim3100$ and a ``late'' data set for the period from day $\sim3100$ through
the final measurements on day 4930. The reason for this splitting of the
data set is that the decline rate $\beta$ is clearly much steeper after day $\sim3100$, which, for illustrative purposes, we have shown in Figures \ref{fig1}, \ref{fig2}, \ref{fig8}, \ref{fig11}, and \ref{fig13} as an exponential with e-folding time of 1100 days. Of course, the transition interval is gradual but, in order to maintain model simplicity, we have taken day 3100 as marking the break between the early and late fitting procedures.

\subsection{Early data fitting}

\begin{figure}
\figurenum{1}
\epsscale{0.8}
\plotone{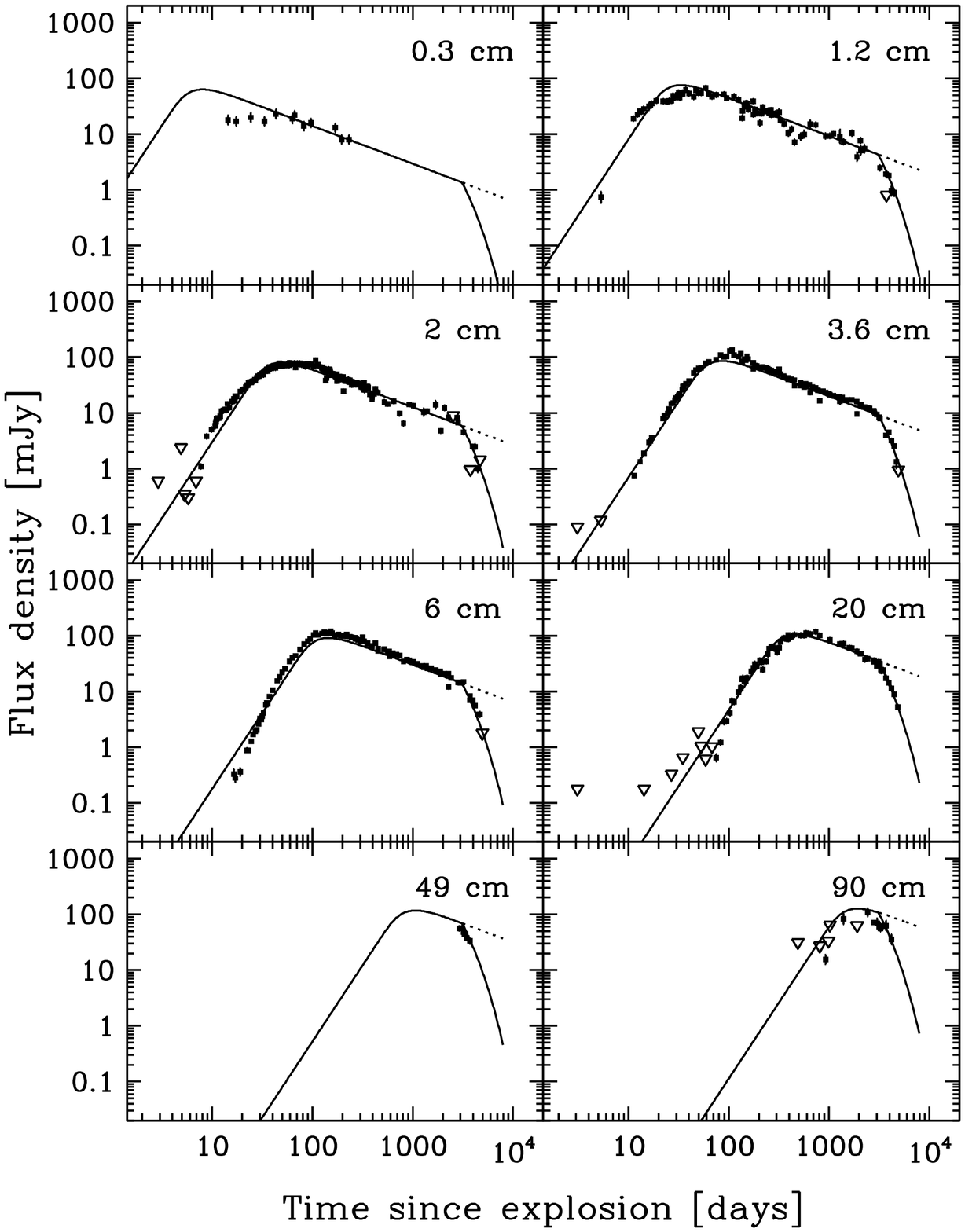}
\figcaption[fig1]{The radio light curves for SN~1993J are plotted from left to right and top to bottom at 0.3, 1.2, 2, 3.6, 6, 20, 49, and 90 cm.  The solid lines represent the best fit ``pure'' synchrotron self-absorption (SSA) model as described in the text with the parameters listed in Table \ref{tab4}, Column 2 and an exponential flux density decline after day 3100 with an e-folding time of 1100 days. The extrapolation of the best-fit model curves without the exponential roll-off is shown as the dashed lines. Upper limits ($3\sigma$) are shown as open inverted triangles ($\triangledown$).
\label{fig1}}
\end{figure}

\begin{figure}
\figurenum{2}
\epsscale{0.8}
\plotone{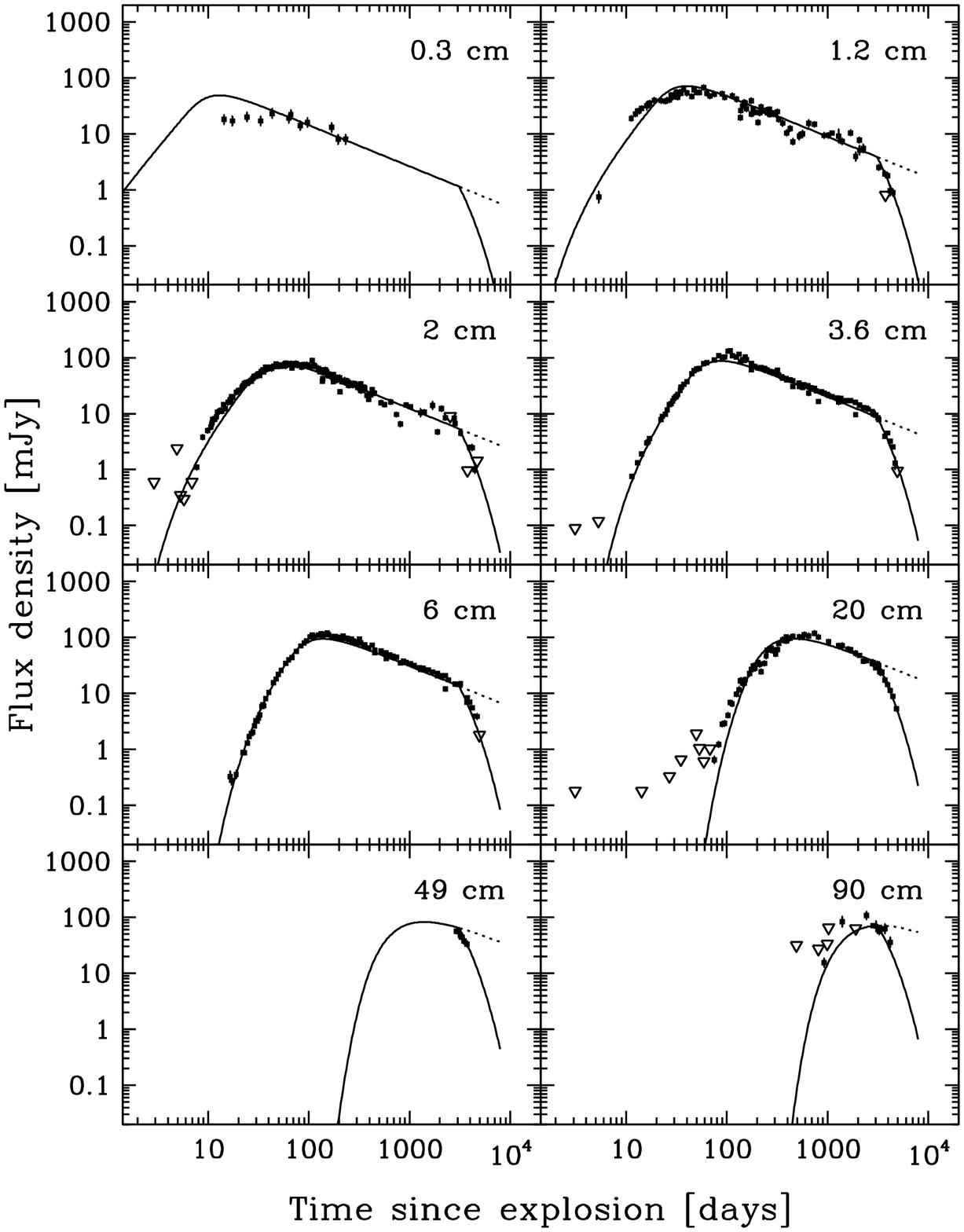}
\figcaption[fig2]{The radio light curves for SN~1993J are plotted from left to right and top to bottom at 0.3, 1.2, 2, 3.6, 6, 20, 49, and 90 cm.  The solid lines represent the best fit ``pure'' thermal, free-free absorption (FFA) model as described in the text with the parameters listed in Table \ref{tab4}, Column 3 and an exponential flux density decline after day 3100 with an e-folding time of 1100 days. The extrapolation of the best-fit model curves without the exponential roll-off is shown as the dashed lines. Upper limits ($3\sigma$) are shown as open inverted triangles ($\triangledown$).
\label{fig2}}
\end{figure}

\begin{figure}
\figurenum{3}
\epsscale{0.8}
\plotone{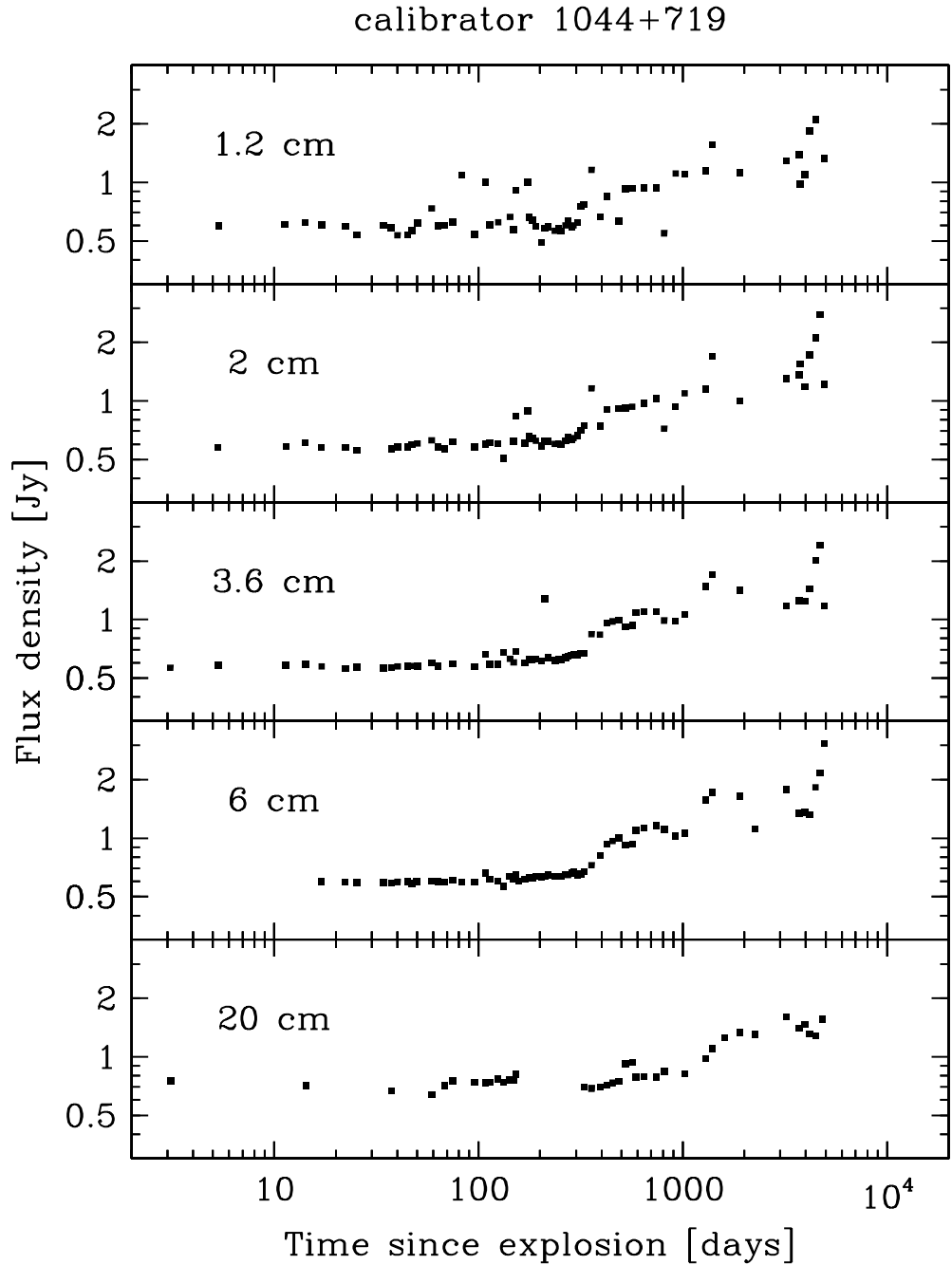}
\figcaption[fig3]{Flux density measurements for the VLA secondary
calibrator J1048+717 at wavelengths of 1.2, 2, 3.6, 6, and 20 cm. Calibration measurements at other observing bands were too sparse to show any trends.
\label{fig3}}
\end{figure}

\begin{figure}
\figurenum{4}
\epsscale{1.0}
\plotone{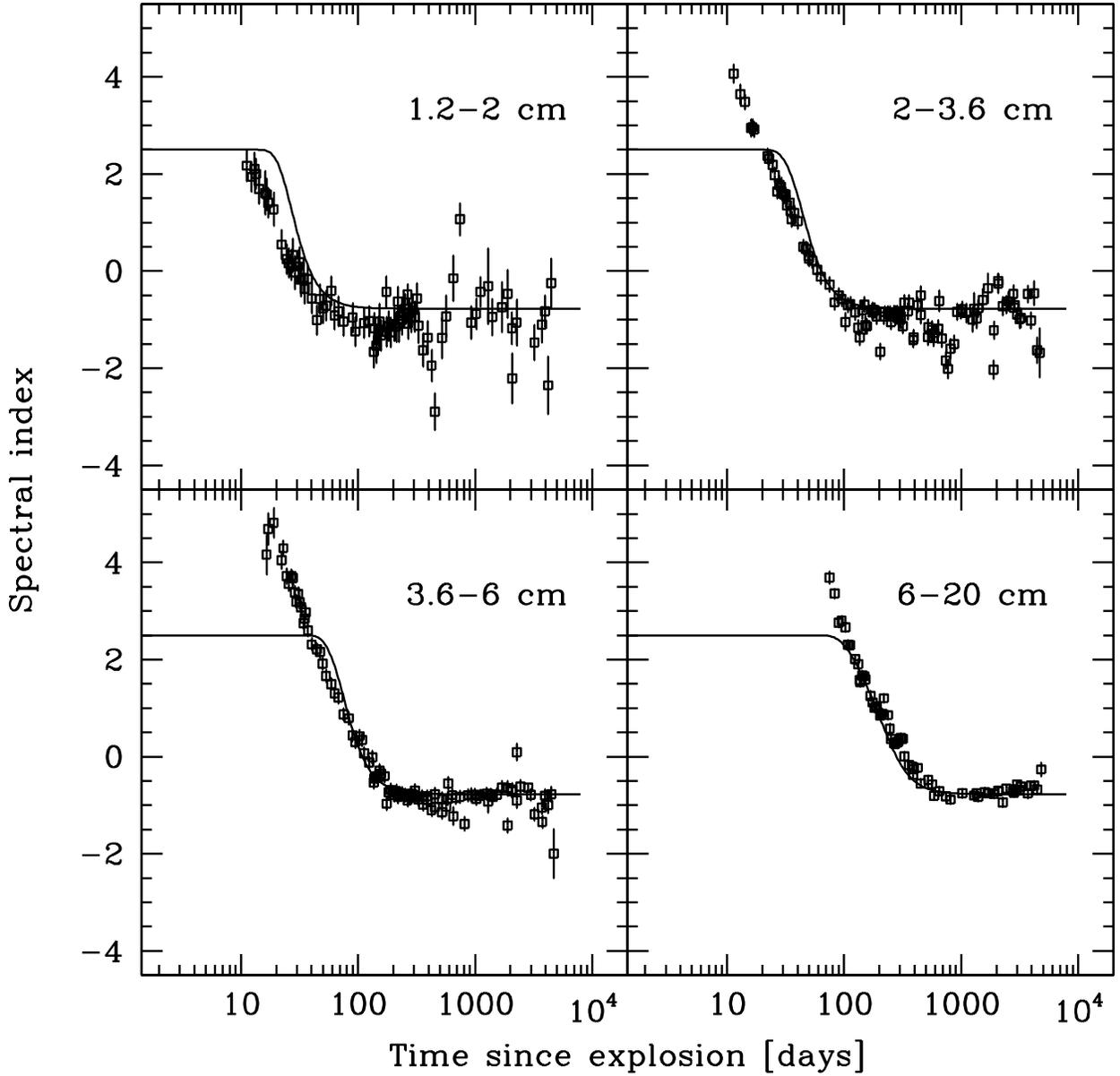}
\figcaption[fig4]{The spectral index ($\alpha$; S $\propto \nu^{+\alpha}$) evolution for SN~1993J between 1.2 and 2 cm (top left), between 2 and 3.6 cm (top right), between 3.6 and 6 cm (bottom left), and between 6 and 20 cm (bottom right).  As in Figure \ref{fig1} the lines represent the best fit pure synchrotron self-absorption (SSA) model as described in the text with the parameters listed in Table \ref{tab4}, Column 2. Note that the observed spectral index values at early times are much in excess of the asymptotic SSA value of $\alpha = +2.5$.
\label{fig4}}
\end{figure}

\begin{figure}
\figurenum{5}
\epsscale{1.0}
\plotone{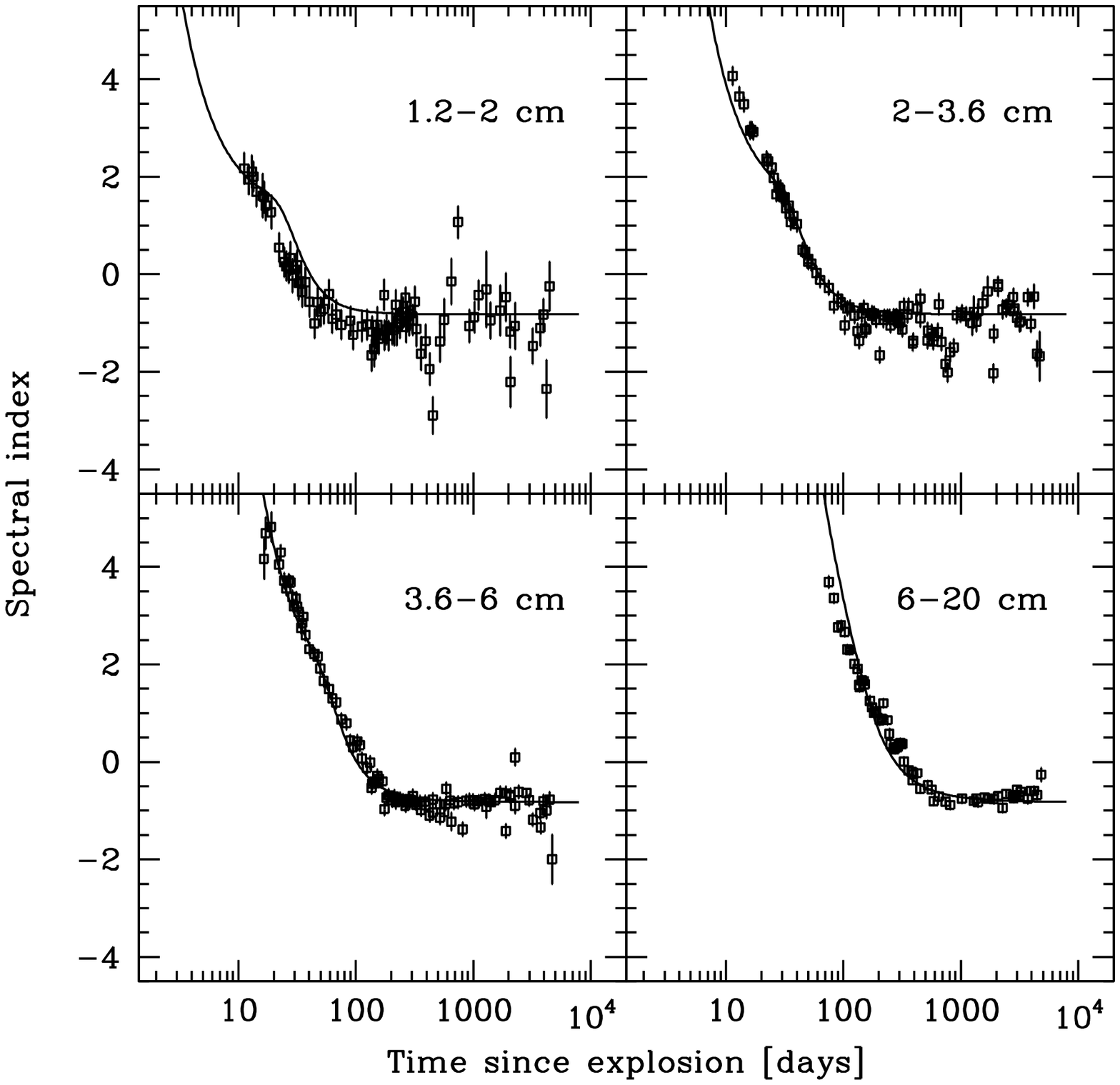}
\figcaption[fig5]{The spectral index ($\alpha$; S $\propto \nu^{+\alpha}$) evolution for SN~1993J between 1.2 and 2 cm (top left), between 2 and 3.6 cm (top right), between 3.6 and 6 cm (bottom left), and between 6 and 20 cm (bottom right).  As in Figure \ref{fig2} the lines represent the best fit pure thermal, free-free absorption (FFA) model as described in the text with the parameters listed in Table \ref{tab4}, Column 3. 
\label{fig5}}
\end{figure}

The early (day $<3100$) data were first fitted with two possible ``pure'' absorption models, namely
pure SSA (i.e. negligible FFA at all times) and pure FFA
(i.e.  negligible SSA at all times).
The  parameters derived from these fits are listed in Table \ref{tab4}, Columns 2 and 3, respectively, 
and the resulting curves are plotted with the data in Figures \ref{fig1}
and \ref{fig2}, respectively.  We also show the spectral index evolution calculated for
pairs of frequencies in Figures \ref{fig4} and \ref{fig5}. The lines in this case are derived from exactly the same models (pure SSA for Figure \ref{fig4} and pure FFA for Figure \ref{fig5}) as were derived for Figures \ref{fig1} and \ref{fig2}. We see that both models are able to represent the light curves as well as the spectral index evolution fairly well. However, there are some features that are
systematically misrepresented by these ``pure'' models, which we discuss below. 

\begin{figure}
\figurenum{6}
\epsscale{0.9}
\plotone{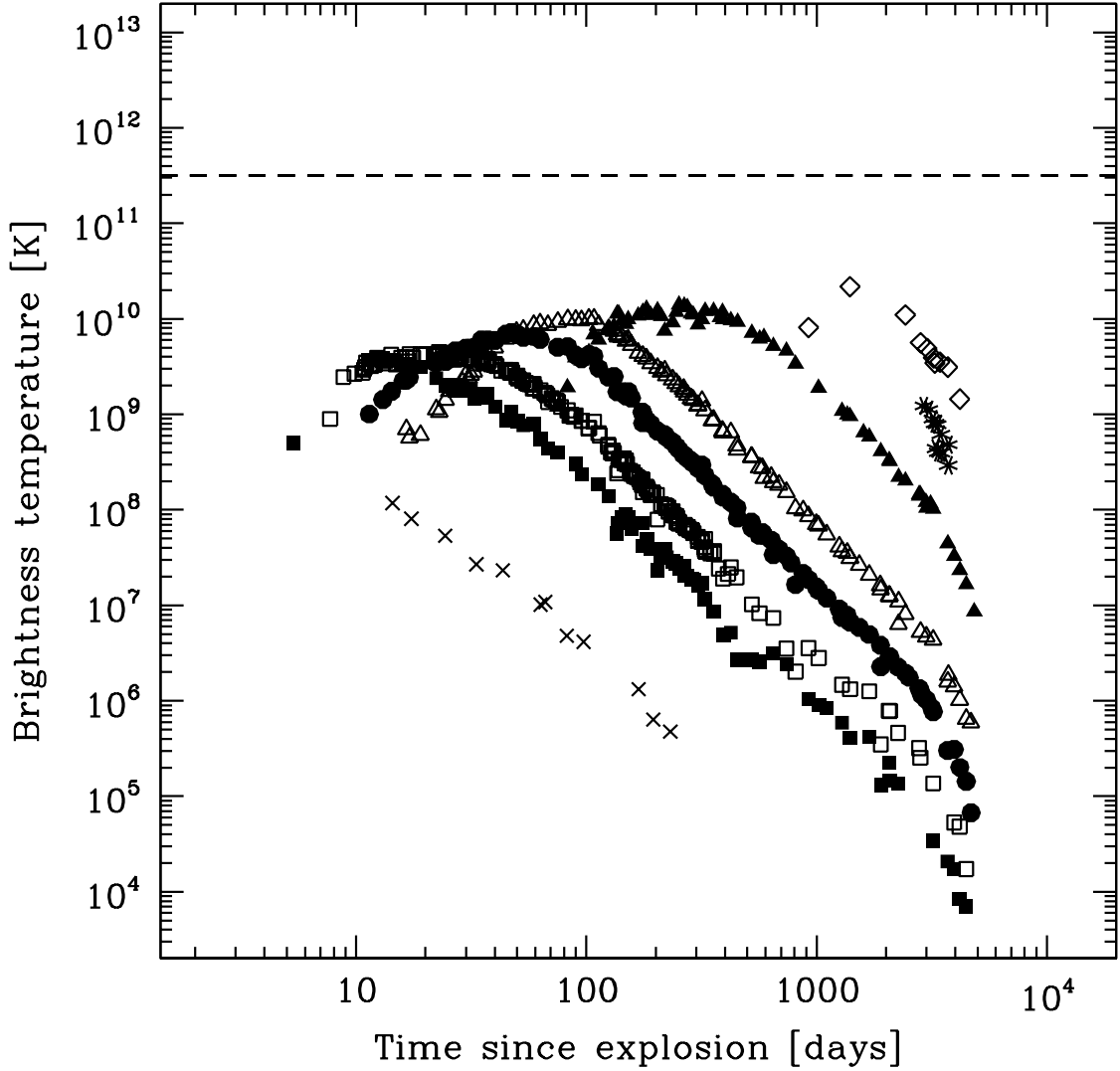}
\figcaption[fig6]{The brightness temperature (T$_B$) evolution for
SN~1993J for, from left to right, 0.3 cm (cross), 1.2 cm (filled square), 2 cm (open square), 3.6 cm (filled circle), 6 cm (open triangle), 20 cm (filled triangle), 49 cm (star), and 90 cm (open diamond) for the case of a pure synchrotron self-absorption (SSA) model as described in the text with the parameters listed in Table \ref{tab4}, Column 2. The horizontal dashed line denotes the limiting value of T$_B\simeq 3\times 10^{11}$K \citep{Kellermann69,Readhead94}, which is not reached at any frequency.  Note that the brightness temperature is low at early times, reaches a peak which always falls well below $3\times 10^{11}$ K, and occurs at later times for lower frequencies.
\label{fig6}}
\end{figure}

\begin{figure}
\figurenum{7}
\epsscale{0.9}
\plotone{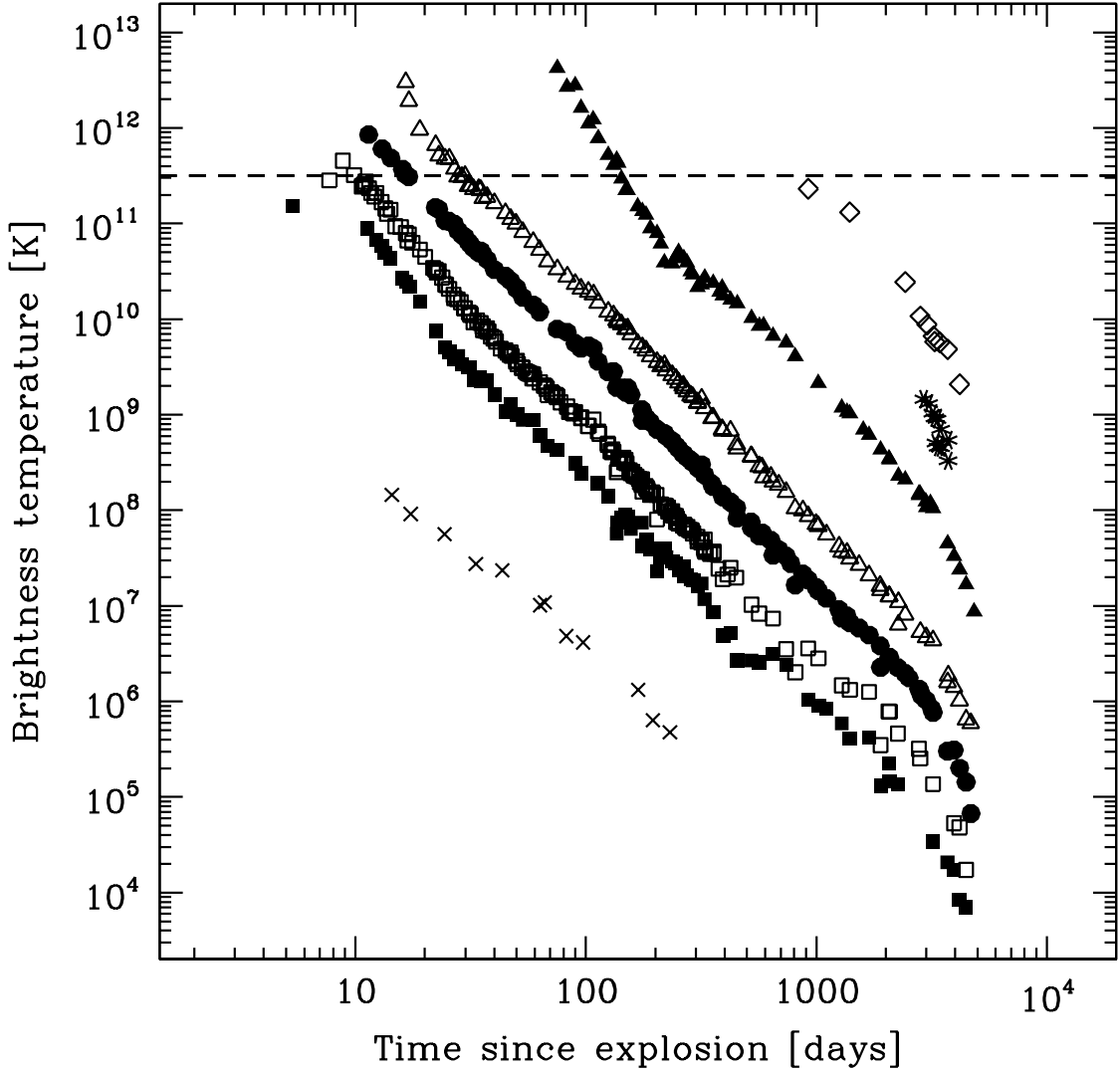}
\figcaption[fig7]{The brightness temperature (T$_B$) evolution for SN~1993J for, from left to right, 0.3 cm (cross), 1.2 cm (filled square), 2 cm (open square), 3.6 cm (filled circle), 6 cm (open triangle), 20 cm (filled triangle), 49 cm (star), and 90 cm (open diamond) for the case of a pure thermal, free-free absorption (FFA) model as described in the text with the parameters listed in Table \ref{tab4}, Column 3. To obtain the ``true'' brightness temperature at early times the measured flux densities have been corrected for the model estimated external, thermal, free-free absorption. The horizontal dashed line denotes the limiting value of T$_B\simeq 3\times 10^{11}$K \citep{Kellermann69,Readhead94}, which is greatly exceeded for most frequencies at early times.
\label{fig7}}
\end{figure}

\begin{figure}
\figurenum{8}
\epsscale{0.8}
\plotone{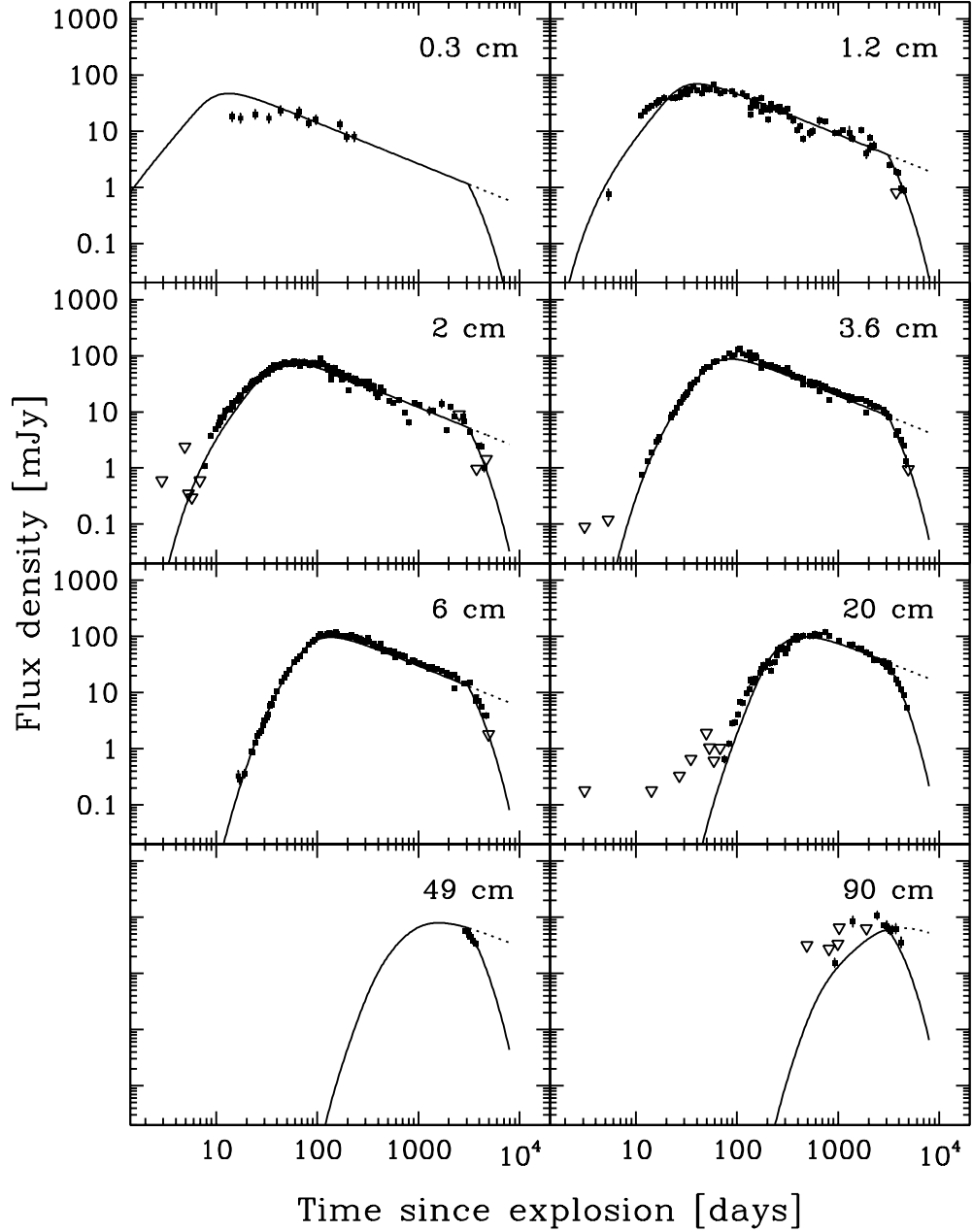}
\figcaption[fig8]{The radio light curves for SN~1993J are plotted from left to right and top to bottom at 0.3, 1.2, 2, 3.6, 6, 20, 49, and 90 cm.  The solid lines represent the best fit combined synchrotron self-absorption (SSA) and thermal, free-free absorption (FFA) model as described in the text with the parameters listed in Table \ref{tab4}, Column 4 and an exponential flux density decline after day 3100 with an e-folding time of 1100 days. The extrapolation of the best-fit model curves without the exponential roll-off is shown as the dashed lines. Upper limits ($3\sigma$) are shown as open inverted triangles ($\triangledown$).
\label{fig8}}
\end{figure}

We also calculate the apparent brightness temperature evolution from Equation \ref{eq13} for both the pure SSA and pure FFA models and plot them in Figures \ref{fig6} and \ref{fig7}, respectively. Note that, for Figures \ref{fig6} and \ref{fig7}, we have applied a correction derived from our models for the presumed external thermal absorption for the early data to obtain a ``true'' flux density, and thus a ``true'' brightness temperature, as if no thermal absorption were present.

For the pure SSA model (Figure \ref{fig1}), the rising branch of the light curve at early times 
tends to be ``too straight'' in that it cannot reproduce the apparent curvature in the flux density turn-on that is most noticeable at frequencies lower than 5 GHz.  Also, the spectral index evolution at early times (Figure \ref{fig4}) is clearly inadequate to represent the observations. This is to be expected because with pure SSA the asymptotic spectral index cannot exceed $\alpha = +2.5$ ($S \propto \nu^{+\alpha}$), clearly at variance with values of $\alpha = 4~{\rm to}~5$ observed at early epochs. Moreover, the corresponding brightness temperature evolution seen in Figure \ref{fig6} is rather strange in that the temperature appears to peak at later times for lower frequencies. 

For the pure FFA model the rising branch of the light curves appears to provide a better fit for both the light curves (Figure \ref{fig2}) and the spectral index (Figure \ref{fig5}), but it fails the test of the brightness temperature (Figure \ref{fig7}). After flux density correction for the external, thermal absorption, the implied brightness temperature at early times, for several frequencies, exceeds the physical limit of $T_B \simeq 3\times 10^{11}$ K \citep[see, \eg][]{Kellermann69,Readhead94}.

Thus, it is clear that no ``simple''  model, either pure SSA or pure FFA, can account for all observational
aspects of the data and a combination of the two absorption mechanisms must be at work.

\begin{figure}
\figurenum{9}
\epsscale{1.0}
\plotone{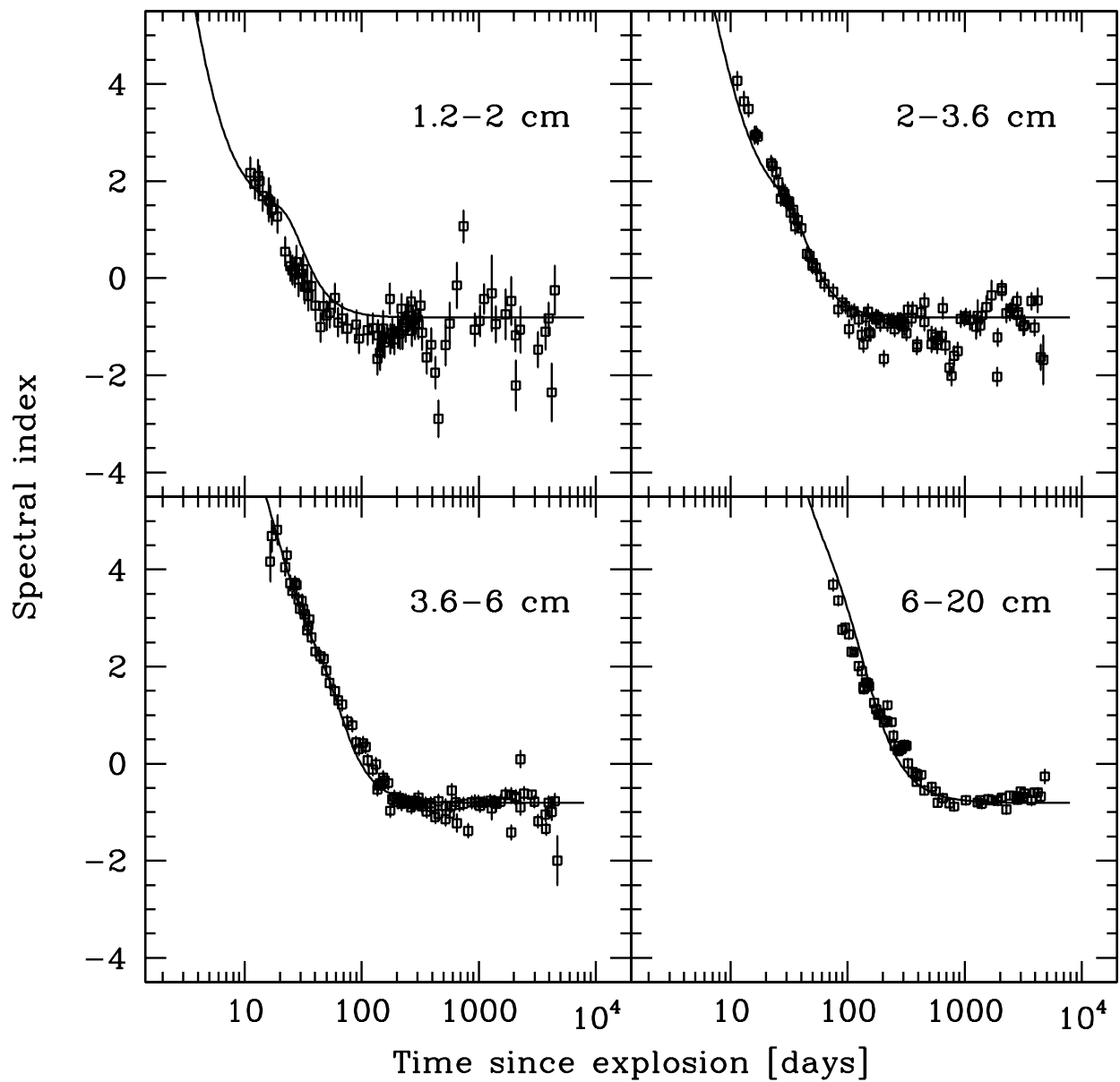}
\figcaption[fig9]{The spectral index ($\alpha$; S $\propto \nu^{+\alpha}$) evolution for SN~1993J between 1.2 and 2 cm (top left), between 2 and 3.6 cm (top right), between 3.6 and 6 cm (bottom left), and between 6 and 20 cm (bottom right).  As in Figure \ref{fig8} the lines represent the best fit combined synchrotron self-absorption (SSA) and thermal, free-free absorption (FFA) model as described in the text with the parameters listed in Table \ref{tab4}, Column 4. 
\label{fig9}}
\end{figure}

\begin{figure}
\figurenum{10}
\epsscale{1.0}
\plotone{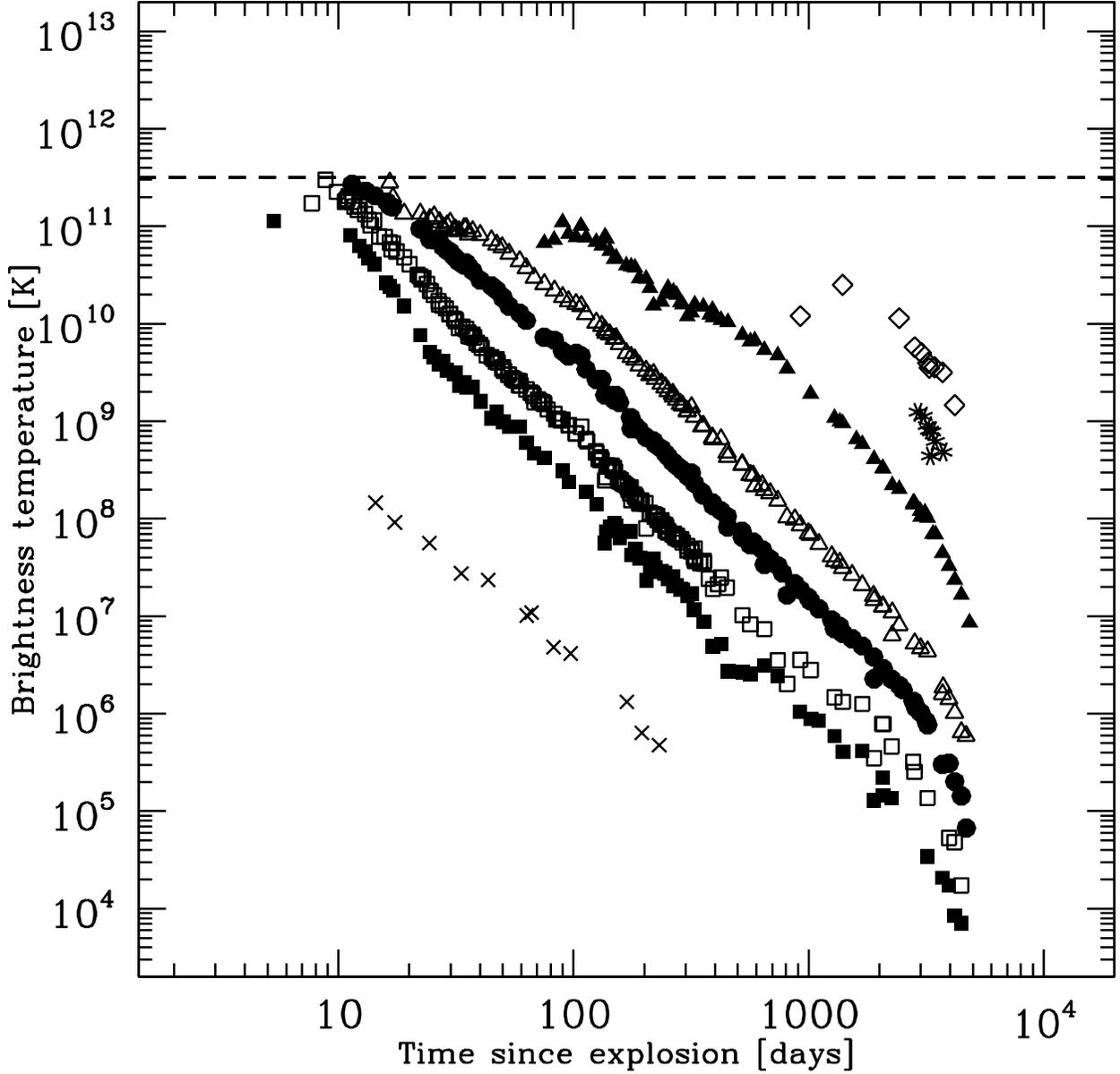}
\figcaption[fig10]{The brightness temperature (T$_B$) evolution for SN~1993J for, from left to right, 0.3 cm (cross), 1.2 cm (filled square), 2 cm (open square), 3.6 cm (filled circle), 6 cm (open triangle), 20 cm (filled triangle), 49 cm (star), and 90 cm (open diamond) corrected for extended, free-free absorption flux density suppression at early times as described in the text, with the parameters listed in Table \ref{tab4}, Column 4. The horizontal dashed line denotes the limiting value of T$_B\simeq 3\times 10^{11}$K \citep{Kellermann69,Readhead94}, which is not exceeded at any frequency.
\label{fig10}}
\end{figure}

The best results are achieved with a model that comprises both SSA and FFA; the best fit parameters are given in Table 4, Column 4 and the corresponding fits are displayed in Figures \ref{fig8}, \ref{fig9}, and \ref{fig10} which show the light curves, the spectral index evolution, and the brightness temperature evolution, respectively. As before, after day $\sim3100$ we have shown, for illustrative purposes, an exponential with an e-folding time of 1100 days for the plotted curve. These results are discussed further in $\S$\ref{Discussion}

\subsection{Late time data fitting}

Even though the data are well described after day $\sim3100$ by an exponential decay with an e-folding time of 1100 days, to show the increasing decline rate of the flux density at all wavelengths, it is perhaps worthwhile to describe the decline after day 3100 in terms of our standard model parameters.

Since all of the absorption processes are negligible by the time
of the steepening flux density decline around day $\sim3100$, the ``late'' radio
light curves are essentially the same for both thermal, free-free absorption (FFA) and 
non-thermal synchrotron-self absorption (SSA) models. Therefore, the only two remaining parameters to be determined are the spectral index ($\alpha$) and the decline rate ($\beta$). Least squares fitting reveals that the ``late'' data are consistent with a constant spectral index equal to that derived for
the early data ($\alpha = -0.8$) but with a much faster rate of decline with $\beta$
steepening from -0.7 to -2.7. However, examination of Figure \ref{fig11} shows that such a fit (shown as the dotted lines) does not describe the data well and an exponential decline with an e-folding time of 1100 days (the solid lines in Figure \ref{fig11}) provides a much better fit. \cite{Chandra04} have proposed a steepening of the spectral index after day 3200 but Figures \ref{fig9} and \ref{fig11}, both of which use constant spectral index models, do not appear to confirm their suggestion (see also $\S$\ref{SI}).

\section{Discussion \label{Discussion}}

\subsection{Synchrotron Self-Absorption vs. Thermal Free-Free
Absorption}

\citet{Chevalier98} proposed that nonthermal synchrotron self-absorption
(SSA) could play a significant role in the early turn-on, absorption-dominated phase of the radio emission from supernovae.  While the
possibility is included in the parameterization of the radio emission
discussed above in Equations \ref{eq2} and \ref{eq10}, actual
observational evidence for the difference in turn-on rate expected
between thermal free-free absorption (FFA) and SSA has been difficult to
establish.  Certainly there are valid physical arguments for expecting
SSA to play a role in the radio emission from RSNe which are radio
luminous at early times.

The source brightness temperature (T$_{\rm B}$) is simply proportional to the source flux density (S), corrected where appropriate for suppression of the flux density by external free-free absorption, divided by the source angular size ($\sim\theta^2$).  T$_{\rm B}$ cannot exceed $3 \times 10^{11}$~K \citep{Kellermann69,Readhead94} without being quenched by inverse-Compton scattering and the consequent SSA. The problems in
determining this relatively straightforward quantity are that the source angular size at very early times cannot be directly observed, even with VLBI techniques; there is likely to be some thermal, ionized, absorbing material surrounding these massive exploding stellar systems giving some
level of initial thermal absorption; current models do not include a start-up engine to predict what the flux density would be without any absorption present; and the velocity of expansion of the radio emitting region may well change during the very early phases of the radio supernova phenomenon. All of these factors could lead to a false estimate of the source size and source flux density at early times, and to an incorrect estimate of the source brightness temperature and the likelihood of SSA dominating.

In some objects such as GRBs, where there is evidence for very little external material to give rise to thermal absorption, and the objects are very compact and very radio luminous, the case for SSA seems clear. For example, the nearby GRB~980425 (SN1998bw), although somewhat ambiguous with \citet{Kulkarni98} claiming evidence for SSA while \citet{Weiler01} showing that FFA gives a fit to the data of equal quality, is probably a good example where SSA is dominant, at least early on.

Several authors have discussed the possibility of SSA being a prominent
absorption mechanism for SN 1993J (\citealt{Fransson98,Perez-Torres01,Bartel02}),
whereas \citet{VanDyk94} were able to describe the early absorption
effects entirely through FFA in a circumstellar medium with a density
profile flatter ($\rho_{CSM} \propto r^{-1.5}$) than the $\rho_{CSM} \propto r^{-2}$ expected for a
constant mass-loss rate, constant velocity, presupernova stellar wind. We have investigated the fitting of both pure SSA (Figures \ref{fig1} and \ref{fig4}) and pure FFA (Figures \ref{fig2} and \ref{fig5}) to the extensive data for SN~1993J and find that both models are acceptable from light curve fitting considerations alone ($\chi_{SSA}^2 = 12.8$, $\chi_{FFA}^2 = 8.8$) with each fitting some parts of the light curves slightly better, and some parts of the data slightly worse, than the other. However, when the additional parameter of the brightness temperature evolution is considered (Figures \ref{fig6} and \ref{fig7}) neither the pure SSA nor the pure FFA models can satisfy all physical  conditions, i.e. to reproduce simultaneously the light curves, the spectral index evolution, and the brightness temperature limit. However, a model which includes both SSA and FFA can account rather well ($\chi_{SSA+FFA}^2 = 8.1$) for the observed radio emission from SN~1993J without violating the brightness temperature limit and provides a good description of the spectral index evolution (see Figures \ref{fig8}, \ref{fig9}, and \ref{fig10}).

\subsection{Evidence for a ``Flatter'' Circumstellar Density Profile}

When \citet{VanDyk94} presented multi-frequency radio observations of SN~1993J for the first eight months of monitoring, they concluded that the CSM surrounding the supernova, which was likely established by the SN
progenitor in the last stages of evolution, consists of: (1) a homogeneous medium (K$_2$) with embedded clumpy or filamentary components (K$_3$), and (2) a CSM with a density profile that is significantly flatter ($\rho_{CSM} \propto {\rm r}^{-1.5}$) than the $\rho_{CSM} \propto {\rm r}^{-2}$ expected for a constant mass-loss rate, constant velocity presupernova stellar wind.  Since the density, and therefore the radio emission, is proportional to the ratio of the mass-loss rate ($\dot M$) to the wind speed ($w$), i.e. ($\dot M/w$), and since the wind speed is unlikely to vary on relatively short time scales, \citet{VanDyk94} estimated that the mass-loss rate from the SN~1993J progenitor system decreased from $\sim 10^{-4}$ M$_{\sun}$ yr$^{-1}$ to $\sim 10^{-5}$ M$_{\sun}$ yr$^{-1}$ during the last 1000 years before explosion.  This conclusion was later supported by \cite{Immler01}, who found a similarly flat CSM density profile from X-ray observations, and accepted by other modeling work \citep{Fransson96}. However, \cite{Fransson98} later concluded that the flatter CSM density profile is not necessary and the results can be interpreted in terms of SSA with an r$^{-2}$ density profile and an electron cooling mechanism. Nevertheless, because the X-ray emission arises from the thermal component of the CSM, rather than the nonthermal component which gives rise to the radio emission, the support for a ``flatter'' CSM density profile from both radio and X-ray observations appears strong.

With a model that includes both SSA and FFA, the radio data alone are not able to constrain the CSM profile efficiently because fits with steep $\delta$ slopes and high values of the K$_2$ can provide comparable accuracy to shallow $\delta$ slopes and low K$_2$ values. Therefore, to calculate our best model, which includes both SSA and FFA and satisfies all observational and physical constraints, we adopted a CSM density profile of $\rho_{CSM} \propto {\rm r}^{-1.61}$, which is close to the behavior determined from the evolution of the X-ray luminosity \citep[][see also $\S$ \ref{Radio-X-Ray}]{Immler01,Zimmermann03}. 

With these assumptions, the mass-loss rate giving rise to uniform, external thermal absorption is given by a straightforward modification of Equation (16) of \citet{Weiler86} which becomes

\begin{eqnarray}
\label{eq14}
\frac{\dot M (M_\odot\ {\rm yr}^{-1})}{( w_{\rm wind} / 10\ {\rm km\
s}^{-1} )} & = & 3.0 \times 10^{-6}\ \phi \ \tau^{0.5} \ m^{-1.5}
{\left(\frac{v_{\rm i}}{10^{4}\ {\rm km\ s}^{-1}}\right)}^{1.5} \times
\nonumber \\ & & {\left(\frac{t_{\rm i}}{45\ {\rm days}}\right) }^{1.5}
{\left(\frac{t}{t_{\rm i}}\right) }^{1.5 m}{\left(\frac{T}{10^{4}\ {\rm
K}} \right)}^{0.68} .
\end{eqnarray}

\noindent Here, the extra factor $\phi$ is a small correction that takes into account the fact that, in this case, the CSM density behaves like $\rho_{CSM} \propto r^{-1.61}$ instead of  $r^{-2}$ as it does under
the usual constant mass-loss assumption. The
factor  $\phi$ is given by the square root of the ratio of the
integration constant for $\tau$  in the case of $\rho_{CSM} \propto 
r^{-1.61}$ to the one appropriate for $\rho_{CSM} \propto  r^{-2}$ , i.e.

\begin{equation}
\label{eq15}
\phi \ = \ \left(\frac{2\times 1.61-1}{2\times 2 - 1}\right)^{0.5} \ = \ 0.86 .
\end{equation}

\noindent For Equation \ref{eq14} we assume $v_{\rm i} = 15,000$ \kms\ at $t_{\rm i} = 45$ days, which is
a value consistent with the results of \cite{Marcaide07} and we adopt values of $T = 20,000$ K, $w_{\rm wind} = 10$ \kms\ (which is appropriate for a RSG wind), and $m=0.845$, as measured  by \cite{Marcaide07}.   With the assumptions for the blastwave and CSM properties discussed above, and the results for the best-fit parameters
listed in Table \ref{tab4}, Column 4, our estimated presupernova mass-loss rate is $\dot M = 5.4\times 10^{-7}\ M_\odot$ yr$^{-1}$ at the time of shock breakout.  

Additionally, we have to take into account the shallow slope of the CSM. The CSM density behaves like $\rho_{CSM} \propto  r^{-1.61}$, indicating that the mass-loss rate was not constant but was higher in the years
leading up to the explosion, i.e. $\dot M \propto r^2 \rho w_{wind} \propto r^{0.39}$ for a constant  $w_{wind}$.   Thus, we calculate that when the abrupt change in the radio light curves occurred around day $\sim3100$ ($\sim8000$ years before explosion) the mass-loss rate was as high as $5.9\times 10^{-6}\ M_\odot$ yr$^{-1}$. Integrating the mass-loss rate over the last $\sim8000$ years, we find that during that time the progenitor star shed $\sim 0.04~M_\odot$ in a massive stellar wind. 

At earlier epochs of the progenitor's evolution, more than 8,000 years before explosion, the mass-loss rate was considerably lower, as indicated by the radio light curve ``break'' discussed above and the transition of the blast wave to a lower density CSM at that time.  One has to keep in mind that these values are derived for an adopted pre-SN stellar wind speed of 10 \kms and blast wave speed of 15,000 \kms.  If the wind speed was appreciably higher than 10 \kms, then the mass-loss rates were proportionally higher.

Thus, our current analysis of this larger data set of radio observations of SN~1993J is consistent with the early predictions of \citet{VanDyk94}, \citet{Immler01}, and \citet{Zimmermann03} of a flatter
CSM density profile and a changing mass-loss rate in the millennia before explosion and inconsistent with a $\rho_{CSM} \propto  r^{-2}$ density profile with electron cooling proposed by \cite{Fransson98}.

\subsection{Increased Flux Density Decline Rate}

\begin{figure}
\figurenum{11}
\epsscale{0.9}
\plotone{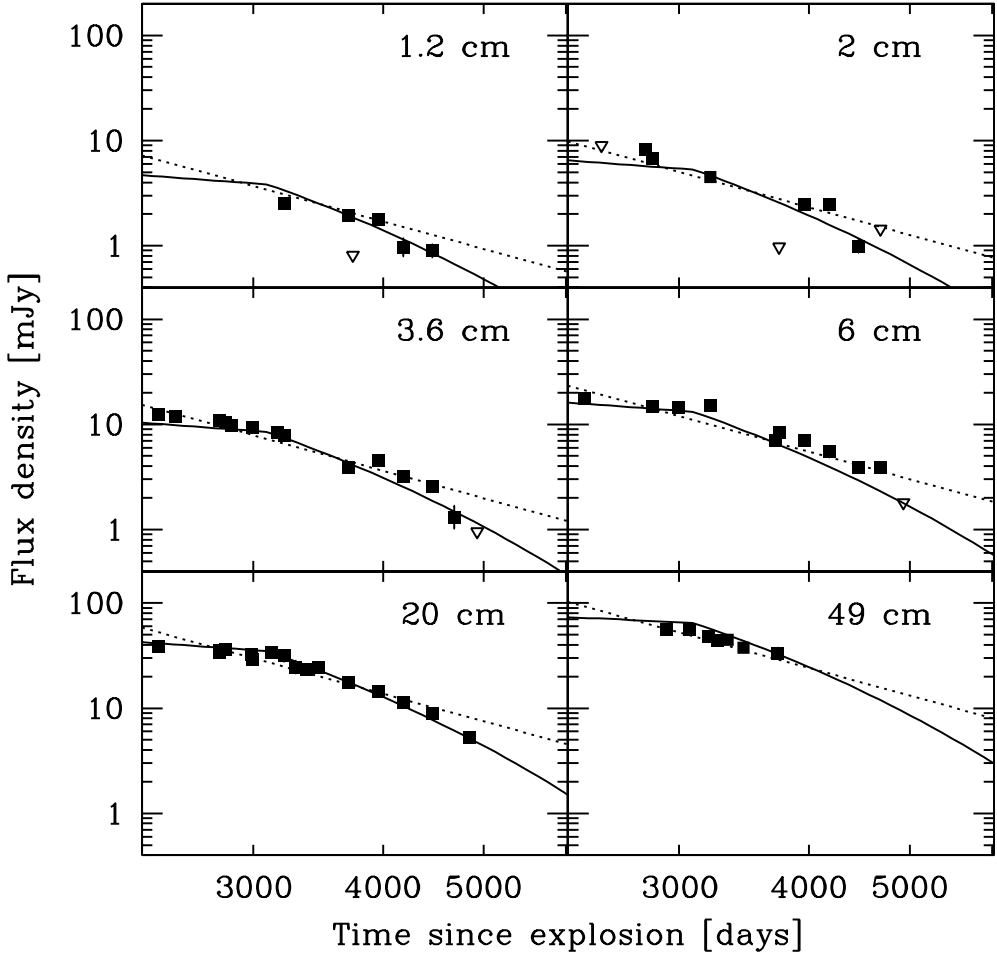}
\figcaption[fig11]{The ``late'' (after day 3100) radio light curves are plotted from left to right and top to bottom at 1.2, 2, 3.6, 6, 20, and 49 cm. Since all absorption processes are negligible, the best model consists of a constant spectral index $\alpha$ fixed from the best fit ``early'' light curve model, a decline rate $\beta$, and a normalization K$_1$.  Whereas the ``early'' data before day 3100 were described by a decline rate of $\beta = -0.7$, the ``late'' data require a decline rate of $\beta = -2.7$, shown as the dotted lines. However, a constant decline rate $\beta$ is clearly not the best description of the data and an exponential decline with an e-folding time of 1100 days (solid lines) is a better description. Upper limits ($3\sigma$) are shown as open inverted triangles ($\triangledown$).
\label{fig11}}
\end{figure}

A noteworthy aspect of the radio emission from SN~1993J is that after
day $\sim3100$, its decline rate significantly steepens. In Figures
\ref{fig1}, \ref{fig2}, \ref{fig8}, \ref{fig11}, and \ref{fig13} this has been illustrated by
multiplying the curve fitted to the early data by an exponential decay term that
affects the emission after  day 3100 and has an e-folding time of 1100
days, i.e. $exp(-(t-t_0-3100)/1100)$ for $t-t_0>3100$.

While the visual description of the data is greatly enhanced by these
curves, another way of describing this change in evolution is shown as
the dotted lines in Figure \ref{fig11} where the data after day 3100 are
fitted with the best fit ``early'' spectral index (the emission is
optically thin at that time, so whether an SSA or FFA model is used at
early times is of no consequence) and a new decline rate $\beta = -2.7$
determined. Nevertheless, it should be noted that the exponential decay
(solid lines in Figure \ref{fig11}) give a better description of the data
than a power-law decline (dotted lines in Figure \ref{fig11}).

Although our exponential decay is a purely empirical assumption, the fact that it is so successful in fitting all light curves at all frequencies simultaneously indicates that the decay is the  result of a
phenomenon whose e-folding time at all frequencies is very short. In other words, the observed decay appears to be dominated by the decline of a synchrotron emission energy supply, such as that derived from the blast wave-CSM interaction, implying a sudden variation in the circumstellar density (and, therefore, in the mass-loss rate) rather than an energy loss reflecting the cooling times at individual frequencies.

\begin{figure}
\figurenum{12}
\includegraphics[angle=-90,scale=.65]{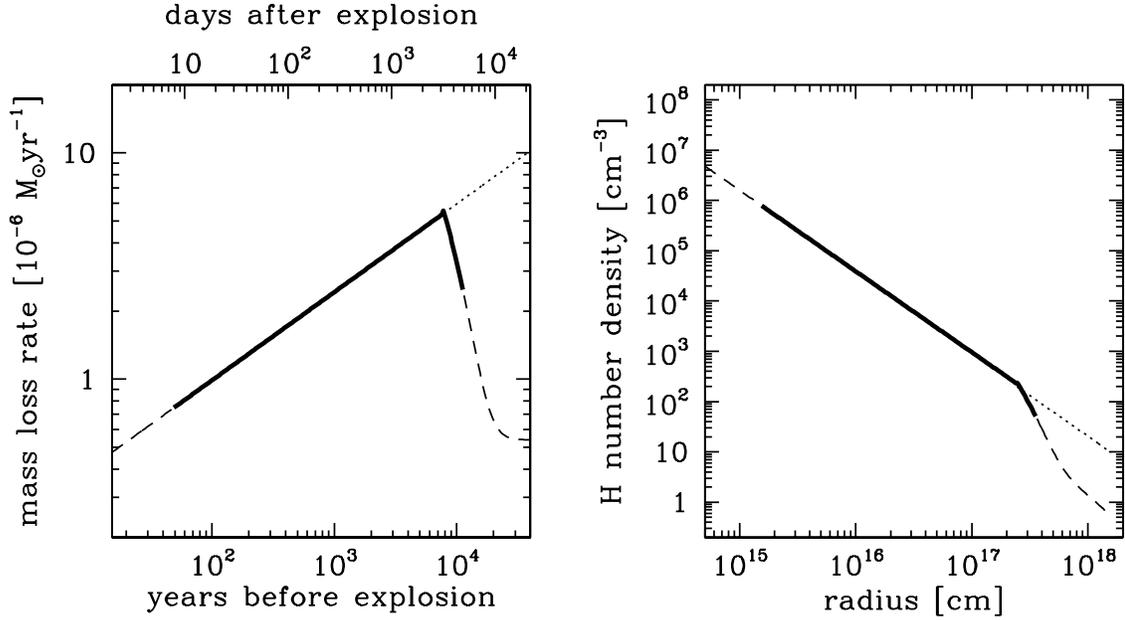}
\figcaption[fig12]{(left) The pre-supernova mass loss rate as function of the time before explosion and (right) the CSM hydrogen number density as a function of radius. The heavy solid curves correspond to behaviors actually constrained by the radio observations, whereas the dashed curves are extrapolations as a simple power law very near to the star and an exponential cutoff plus a constant mass-loss rate at large times (radii) before explosion. This last, an assumed constant mass-loss rate at large times before explosion is simply notional since our observations provide no constraints at such times (radii). The dotted lines are power law extrapolations of the density for larger radii or the mass-loss rate at earlier epochs, which are drawn just to guide the eye to better appreciate the variations. \label{fig12}}
\end{figure}

\begin{figure}
\figurenum{13}
\epsscale{0.7}
\plotone{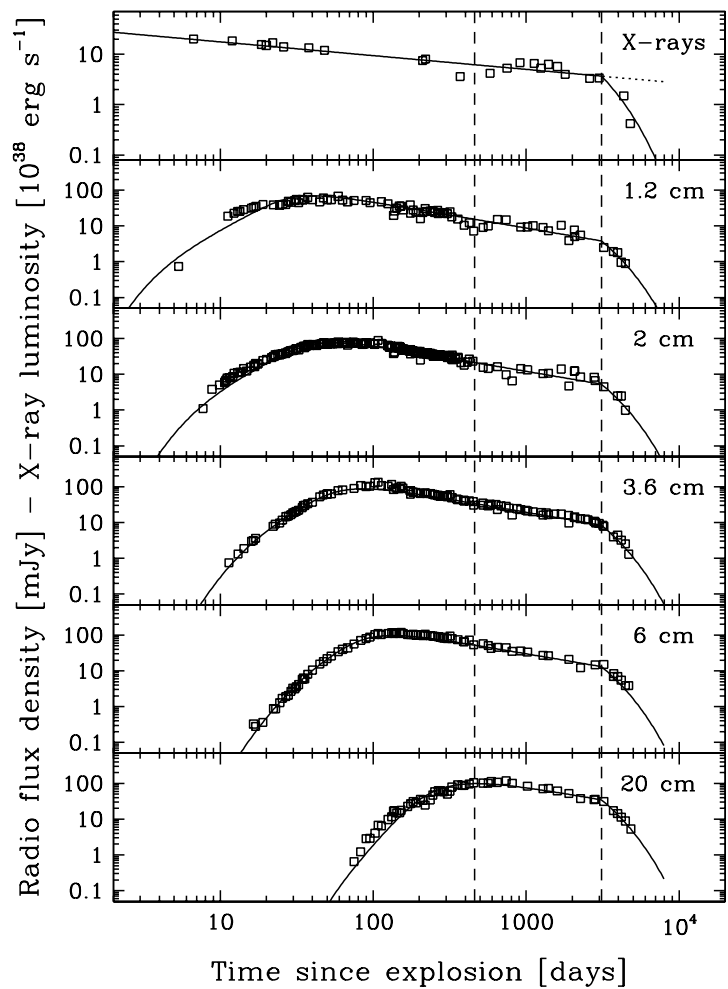}
\figcaption[fig13]{The X-ray and the best populated radio data sets (no upper limits are shown) are plotted on the same time scale for comparison. Each frame is labeled with the wavelength of the observations. The vertical dashed lines denote the epoch of a prominent dip in the X-ray light curve (around day $\sim460$) and the beginning of the overall decay at all frequencies (around day $\sim3100$).  Note that the first X-ray dip corresponds to a similar dip in the 1.2 cm radio light curve that is not prominent at longer radio wavelengths. \label{fig13}}
\end{figure}

Interpreting this exponential decay in terms of a sudden decrease of the CSM density leads to a pre-supernova density distribution that decreases like $r^{-1.61}$ up a distance of  2.4$\times 10^{17}$cm, and has a sudden drop by a factor of $\sim3$ by a radius of $\sim 4\times 10^{17}$cm. To cause this, the mass-loss rate had to have a steep enhancement by at least a factor of 3 around 8,000 years before the supernova explosion and to decrease afterwards at a rate proportional to $t^{-0.39}$. This scenario is illustrated in Figure \ref{fig12}, that shows the mass-loss rate as function of the time before explosion (left hand panel) and the H number density as a function of radius (right hand panel). The heavy solid curves correspond to behaviors actually constrained by the radio observations, whereas the dashed curves are extrapolations as a simple power law very near to the star and as an exponential cutoff plus a constant mass-loss rate at large times (radii) before explosion. This last, an assumed constant mass-loss rate at large times before explosion is simply notional since our observations provide no constraints at such times (radii). The dotted lines are power law extrapolations of the density for larger radii or the mass-loss rate at earlier epochs, which are drawn just to guide the eye to better appreciate the variations. 

In astrophysical terms our empirical result suggests that the progenitor star underwent a shell ejection that, $\sim8,000$ years before the supernova explosion, increasing the effective mass-loss rate from the star
by possibly an order-of-magnitude, which then slowly decreased with time. This phenomenon is reminiscent of the recurrent shell ejections considered by Panagia and Bono (2001) for stars of masses around 12-14
M$_\odot$ that become pulsationally unstable in their red supergiant
phases.

\subsection{Spectral Index Evolution \label{SI}}

In addition to the radio light curves and their comparison with models, it is also possible to examine the spectral index evolution and its comparison with pure SSA (Figure \ref{fig4}) and pure FFA (Figure
\ref{fig5}) model predictions. Examination of both figures shows that the spectral index evolution is reasonably well described only by the FFA model. However, as discussed earlier, the pure FFA model appears unrealistic when brightness temperature considerations (Figure \ref{fig7}) are included. Our best model, a combination of both SSA and FFA processes (Figure \ref{fig8}), fits the spectral index evolution (Figure \ref{fig9}) quite well and satisfies the brightness temperature limitations (Figure \ref{fig10}). Nevertheless, deviations can still be noticed in the light curves and in the spectral index evolution because, as mentioned earlier, we cannot expect simple models to describe such a complex phenomenon as a supernova explosion in detail.

Particular note should be made of the late time spectral index evolution. \cite{Chandra04} suggest a break in the spectrum around day $\sim3200$, with higher frequency flux densities declining faster than lower frequency ones, leading to a steepening of the spectral index. From this they were able to calculate a number of physical properties of the radio emitting region. With data now extending to day $\sim5000$ it is possible to check for such a break. Examination of the agreement between our constant spectral index model curves and the data, particularly in Figure \ref{fig11}, does not appear to confirm this suggestion. 

\subsection{Radio and X-ray evolution \label{Radio-X-Ray}}

A comparison of the radio and X-ray light curves reveals features that can help to understand the SN~1993J
phenomenon.  The top panel of Figure \ref{fig13} shows the X-ray light curve as summarized by \cite{Zimmermann03} plus two recent measurements obtained with the SWIFT satellite (S. Immler, private communication). We have calculated a fit to the upper envelope of the X-ray data that gives $L_X \propto t^{-0.22}$, i.e. a slope that is a little flatter than the \citet{Zimmermann03} value of -0.30, and that implies a density behavior of $\rho_{CSM} \propto r^{-1.61}$, again marginally flatter than \citet{Zimmermann03} estimate of $\rho_{CSM} \propto r^{-1.65}$.

The other five panels display the observed radio data (without showing
upper limits) at 1.2, 2, 3.6, 6, and 20 cm and their best-fit curves as
already shown in Figure \ref{fig8} and given in Table \ref{tab4},
Column 4. The two vertical dashed lines are meant to guide the eye to
two particular events, the right-most being the already discussed
steepening of the flux density decline rate of the radio emission after
day $\sim3100$, and the left-most being an apparent ``dip'' of the X-ray
luminosity around day $\sim460$. Even if the X-ray coverage at late
times is rather sparse, the two most recent measurements  appear to
confirm a  steepening seen in the radio decay rate after day
3100, consistent with the same e-folding time of 1,100 days. Also, it is intriguing that the radio light curve at 1.2 cm
appears to ``dip'' in a manner similar to the X-ray luminosity around
day $\sim460$. Both the X-ray and the 1.2 cm light curves are well
enough sampled to provide rough time scales for this event; i.e., the
dip has $\sim70$ days half-power width at 1.2 cm and $\sim280$ days
half-power width at X-ray wavelengths. 

Since radio emission is due to synchrotron processes, whereas the X-ray
emission is accounted for by reverse shock heating, these coincidences
in the overall evolution suggest that the observed variations are the
result of a change in the efficiency of the energy supply, \ie most
likely due to anomalies in the circumstellar medium density
distribution. 

On the other hand, it is not clear why a significant variation is
observed in the 1.2 cm flux density at the time of the  X-ray dip around
460 days, but no appreciable changes are recorded at other equally well 
sampled radio frequencies.

\section{Conclusions}

We present detailed radio observations of SN~1993J at multiple wavelengths for $\sim13$ years after explosion. This data collection arguably represents the most detailed set of observations of any supernova at any wavelength except for the nearby and spectacular SN~1987A.

The radio emission evolves regularly in both time and frequency, and the usual interpretation in terms of shock interaction with a complex circumstellar medium (CSM) formed by a pre-supernova structured stellar wind, with the inclusion of both synchrotron self-absorption (SSA) and thermal free-free absorption (FFA) at early times, describes the observations rather well considering the complexity of the phenomenon.  However, there are some notable characteristics peculiar to SN~1993J. 1) At a time around day $\sim3100$ after shock breakout the decline rate of the radio emission steepens from (t$^{+\beta}$) $\beta \sim -0.7$ to $\beta \sim -2.7$ without change in the spectral index ($\nu^{+\alpha}$; $\alpha \sim -0.81$). This variation, however, can better be described in terms of an exponential decay starting at day $\sim3100$ with an e-folding time of $\sim 1100$ days. 2) The spectral index appears constant throughout our measurement era. 3) The best overall fit to all of the ``early'' (\i.e., before day 3100) data is a model including  both SSA and FFA components, evolving to a constant decline rate until the break at day $\sim3100$. In particular, neither a pure SSA nor a pure FFA absorbing model can provide a  fit that simultaneously reproduces the light curves and the spectral index evolution and provides a physically realistic brightness temperature evolution. 4) The radio and X-ray light curves display quite similar behavior and their comparison suggests the presence of at least two episodes of change in the supernova progenitor mass-loss rate in the last several thousand years before explosion.

\acknowledgements
We are indebted to the VLA TAC and schedulers for permitting and
arranging our numerous observations over many years and to observers who have contributed
data at other radio wavelengths, sometimes unpublished. KWW wishes to
thank the Office of Naval Research (ONR) for the 6.1 funding supporting
his research.  CJS is a Cottrell Scholar of Research Corporation and work on this project has been supported by the NASA Wisconsin Space Grant Consortium. NP is Astronomer Emeritus at the Space Telescope Science Institute (STScI) that kindly provided research facilities and partial support for this work. JMM acknowledges support from grant AYA2006-14986-C02-02. Additional information and data on radio supernovae can be found on {\it http://rsd-www.nrl.navy.mil/7213/weiler/sne-home.html}
and linked pages.

\clearpage

%Table 1
\begin{deluxetable}{cccccccc}
\tabletypesize{\scriptsize}
\tablewidth{7.0in}
%\table columns(8}
\tablecaption{Flux Density Measurements for SN~1993J\label{tab1}
}
\tablehead{
\colhead{Obs.} & \colhead{Days from} & \colhead{Tel/VLA} & \colhead{S(20cm) $\pm$ err\tablenotemark{a}} & \colhead{S(6cm) $\pm$ err\tablenotemark{a}} & \colhead{S(3.6cm) $\pm$ err\tablenotemark{a}} & \colhead{S(2cm) $\pm$ err\tablenotemark{a}} & \colhead{S(1.2cm) $\pm$ err\tablenotemark{a}} \\
\colhead{Date} &  \colhead{Reference Date} & \colhead{Config.\tablenotemark{b}} & \colhead{(mJy)} & \colhead {(mJy)}  & \colhead{(mJy)} & \colhead{(mJy)} & \colhead{(mJy)}
} 
\startdata
28-Mar-93 &  $= 0.00$  & & & & & & \\
30-Mar-93 &   2.90 & Camb    & \nodata               & \nodata               & \nodata               & $<   0.6$           & \nodata               \\
31-Mar-93 &   3.10 & VLA-B   & \nodata               & \nodata               & $<   0.090$           & \nodata               & \nodata               \\
31-Mar-93 &   3.10 & VLA-B   & $<   0.180$           & \nodata               & \nodata               & \nodata               & \nodata               \\
01-Apr-93 &   4.91 & Camb    & \nodata               & \nodata               & \nodata               & $<   2.4$           & \nodata               \\
02-Apr-93 &   5.29 & VLA-B   & \nodata               & \nodata               & \nodata               & $<   0.348$           & \nodata               \\
02-Apr-93 &   5.30 & VLA-B   & \nodata               & \nodata               & $<   0.120$           & \nodata               & \nodata               \\
02-Apr-93 &   5.34 & VLA-B   & \nodata               & \nodata               & \nodata               & \nodata               & $   0.740\pm    0.195$\\
\enddata  
\tablenotetext{a}{All upper limits are three times the map rms ($3\sigma$).}
\tablenotetext{b}{The ``Camb'' data are all from the Ryle Telescope \citep[and private communication]{Pooley93a,Pooley93b}. For a discussion of why 10\% errors were assumed for the Cambridge data, see $\S$\ref{observations}.}
\tablenotetext{c}{\citet{Bartel02}}
\tablenotetext{d}{\citet{Chandra04}}
\tablenotetext{e}{\citet{PerezTorres02}}
\tablenotetext{f}{\citet{PerezTorres02} give two measurements of SN1993J in the 20~cm band, a value of $30.5\pm 0.4$ at 1.67~GHz and $33.7\pm 0.4$ at 1.34~GHz. We include both measurements here.}
\end{deluxetable}

\clearpage

%Table 2
\begin{deluxetable}{cccccc}
\tabletypesize{\scriptsize}
\tablecaption{Other Flux Density Measurements for SN~1993J\label{tab2}}
\tablewidth{0pt}
\tablehead{
\colhead{Obs.} & \colhead{Days from} 
& \colhead{Telescope} & \colhead{Flux Density $\pm$ Error} & \colhead{Frequency} & \colhead{Ref.\tablenotemark{a}}\\
\colhead{Date} & \colhead{Reference Date} & \colhead{} &
\colhead{} & \colhead{} \\
\colhead{} & \colhead{} & \colhead{} & \colhead{(mJy)} & \colhead{(GHz)} }
\startdata
28-Mar-93 &= 0.00  & & & \\
09-Apr-93 &  12.00 & IRAM   &$ 11.0 \pm  3.0 $ & 87 &1\\
11-Apr-93 &  14.00 & IRAM   &$ 11.0 \pm  3.0 $ & 87 &1\\
11-Apr-93 &  14.39 & OVRO   &$ 18 \pm  4 $ & 99.4 &2\\
12-Apr-93 &  15.00 & IRAM   &$ 13.5 \pm  3.4 $ & 87 &1\\
14-Apr-93 &  17.39 & OVRO   &$ 17 \pm  4 $ & 99.4 &2\\
15-Apr-93 &  18.00 & IRAM   &$ 12.5 \pm  3.2 $ & 87 &1\\
16-Apr-93 &  19.00 & IRAM   &$ 10.0 \pm  2.8 $ &110 &1\\
\enddata
\tablenotetext{a}{References: 1. \cite{Radford93};
  2. \cite{Phillips93b}; 3. S. Radford, private communication;
  4. W. Reich, private communication; 5. \cite{Phillips93c};
  6. Current paper; 7. \cite{Bartel02}; 8. \cite{Perez-Torres01}; 9. \cite{Chandra04}}
\tablenotetext{b}{Calibrator is J1331+305 (3C286)}
\end{deluxetable}
	  
\clearpage

%Table 3
\begin{deluxetable}{cccccccc}
\tabletypesize{\scriptsize}
\tablewidth{5.5in}
%\table columns(8}
\tablecaption{Flux Density Measurements for the VLA Secondary Calibrator J1048+717\tablenotemark{a}\label{tab3}
}
\tablehead{
\colhead{Obs.} & \colhead{Days from} & \colhead{Tel/VLA} & \colhead{S(20cm)} & \colhead{S(6cm)} & \colhead{S(3.6cm)} & \colhead{S(2cm)} & \colhead{S(1.2cm)} \\
\colhead{Date} &  \colhead{Reference Date} & \colhead{Config.} & \colhead{(mJy)} & \colhead {(mJy)}  & \colhead{(mJy)} & \colhead{(mJy)} & \colhead{(mJy)}
} 
\startdata
31-Mar-93 &   3.10 & VLA-B   &  \nodata  & \nodata               & $   0.566$            & \nodata               & \nodata               \\
31-Mar-93 &   3.10 & VLA-B   & $   0.754$  & \nodata               & \nodata               & \nodata               & \nodata               \\
02-Apr-93 &   5.29 & VLA-B   & \nodata               & \nodata               & \nodata               & $   0.577$            & \nodata               \\
02-Apr-93 &   5.30 & VLA-B   & \nodata               & \nodata               & $   0.583$            & \nodata               & \nodata               \\
02-Apr-93 &   5.34 & VLA-B   & \nodata               & \nodata               & \nodata               & \nodata               & $   0.596$            \\
08-Apr-93 &  11.37 & VLA-B   & \nodata               & \nodata               & $   0.582$            & $   0.583$            & $   0.611$            \\
09-Apr-93 &  12.33 & VLA-B   & \nodata               & \nodata               & \nodata               & $   0.591$\tablenotemark{b}            & $   0.616$\tablenotemark{b}            \\
10-Apr-93 &  13.04 & VLA-B   & \nodata               & \nodata               & $   0.585$\tablenotemark{b}            & \nodata               & $   0.619$\tablenotemark{b}            \\
\enddata
\tablenotetext{a}{For the measurements from 11 September 1993 through 08 February 1994, the secondary calibrator at 20 cm was J0949+662.} 
\tablenotetext{b}{A primary calibrator was not measured, so the flux density for the secondary calibrator was determined from earlier observations.} 
\end{deluxetable}

\clearpage

%Table 4
\begin{deluxetable}{cccc}
\tabletypesize{\footnotesize}
\tablewidth{4.6in}
\tablecaption{Model Fitting Results for SN~1993J\label{tab4}}
\tablehead{\colhead{Parameter\tablenotemark{a}} & \multicolumn{3}{c}{Early Data\tablenotemark{b}}\\
\cline{2-4} 
\colhead{ }& \colhead{SSA only} &\colhead{FFA only}&\colhead{SSA + FFA}\\
\colhead{ }& \colhead{Fit} &\colhead{Fit}&\colhead{``Best'' Fit}\\
} 
\startdata
$K_1$               & $3.3 \times 10^3$ & $4.9 \times 10^3$ & $4.8 \times 10^3$ \\  
$\alpha$            & -0.77 & -0.82 & -0.81 \\ 
$\beta$             & -0.68 & -0.73 & -0.73 \\ 
$K_2$               & 0\tablenotemark{c} & $1.7 \times 10^2$ & $1.6 \times 10^2$ \\
$\delta$            & \nodata & -1.42 & -1.88 \\
$K_3$               & 0\tablenotemark{c} & $4.3 \times 10^5$ & $4.6 \times 10^5$ \\
${\delta}^{\prime}$ & \nodata & -2.84 & -2.83 \\
$K_4$               & 0\tablenotemark{c} & 0\tablenotemark{c} & 0\tablenotemark{c} \\
\enddata 
\tablenotetext{a}{See the text for an explanation of the model fitting parameters.}
\tablenotetext{b}{Using data from the first radio detection to day 3100.}
\tablenotetext{c}{Defined fixed for the fit.}
\tablenotetext{d}{From \citet{Wheeler93}}
\tablenotetext{e}{From \citet{Freedman94}}
\tablenotetext{f}{Assuming $w_{\rm wind}= 10~{\rm km}~{\rm s}^{-1}$, ${\rm t_i} = 45~{\rm days}$, ${\rm v_{i}} = 15,000~\rm{km}~\rm{s}^{-1}$, 
$\rm{T} = 20,000~\rm{K}$, and ${\rm m = 0.845}$ (see Equation \ref{eq14}).}
\end{deluxetable}

\clearpage

\clearpage

\clearpage

\clearpage

\clearpage

\clearpage

\clearpage

\clearpage

\clearpage

\clearpage

\clearpage

\clearpage

\clearpage

\end{document}